\documentclass{aastex631}

\usepackage{multirow}
\usepackage{booktabs, makecell}
\usepackage{amsmath}

\usepackage{xcolor}
\def\r{\textcolor{red}}
\def\b{\textcolor{blue}}
\def\p{\textcolor{purple}}
\def\w{\textcolor{brown}}
\def\g{\textcolor{green}}

\begin{document}

\title{X-ray Emission Study of TeV Blazars}

\author[0009-0000-4102-9115]{Zhihao Ouyang}
\affiliation{Shanghai Key Lab for Astrophysics, Shanghai Normal University, Shanghai 200234, People's Republic of China}
\affiliation{Center for Astrophysics, Guangzhou University, Guangzhou 510006, People's Republic of China}

\author[0000-0001-8244-1229]{Hubing Xiao}
\affiliation{Shanghai Key Lab for Astrophysics, Shanghai Normal University, Shanghai 200234, People's Republic of China}

\author{Shangchun Xie}
\affiliation{Shanghai Key Lab for Astrophysics, Shanghai Normal University, Shanghai 200234, People's Republic of China}

\author[0009-0008-2290-4349]{Jingyu Wu}
\affiliation{Shanghai Key Lab for Astrophysics, Shanghai Normal University, Shanghai 200234, People's Republic of China}

\author[0000-0003-2551-0202]{Xuhong Ye}
\affiliation{Center for Astrophysics, Guangzhou University, Guangzhou 510006, People's Republic of China}
\affiliation{Key Laboratory for Astronomical Observation and Technology of Guangzhou, Guangzhou 510006, People's Republic of China}
\affiliation{Astronomy Science and Technology Research Laboratory of Department of Education of Guangdong Province, Guangzhou 510006, People's Republic of China}

\author{Junhang Liu}
\affiliation{Shanghai Key Lab for Astrophysics, Shanghai Normal University, Shanghai 200234, People's Republic of China}

\author[0000-0001-7416-7434]{Lili Yang}
\affiliation{School of Physics and Astronomy, Sun Yat-Sen University, Zhuhai 519082, People’s Republic of China}

\author[0000-0003-1530-3031]{Marina Manganaro}
\affiliation{Department of Physics, University of Rijeka, Rijeka 51000, Croatia}

\author{Jianzhen Chen}
\affiliation{Shanghai Key Lab for Astrophysics, Shanghai Normal University, Shanghai 200234, People's Republic of China}

\author[0000-0002-5929-0968]{Junhui Fan}
\affiliation{Center for Astrophysics, Guangzhou University, Guangzhou 510006, People's Republic of China}
\affiliation{Key Laboratory for Astronomical Observation and Technology of Guangzhou, Guangzhou 510006, People's Republic of China}
\affiliation{Astronomy Science and Technology Research Laboratory of Department of Education of Guangdong Province, Guangzhou 510006, People's Republic of China}

\correspondingauthor{Hubing Xiao, Junhui Fan}
\email{hubing.xiao@shnu.edu.cn, fjh@gzhu.edu.cn}




\begin{abstract}

In this work, we present a systematic X-ray study of 85 TeV-detected blazars based on archival \emph{Swift}-XRT and \emph{NuSTAR} observations available up to May 2024. 
We derive unabsorbed X-ray fluxes and photon indices through spectral fitting, and investigate their light curves, spectral energy distributions, flux distributions, power spectral densities, and flux–index correlations.
Using the fractional variability method, we identify 78 sources as variable in the soft X-ray band and 14 sources as variable in the hard X-ray band. 
For well-sampled soft X-ray light curves, the flux distributions are generally better described by log-normal functions, while the PSD slopes span $\beta_{\rm PSD}=0.1$--$1.6$, with most values between 1.0 and 1.6, where the PSD is defined as $P(f)\propto f^{-\beta_{\rm PSD}}$.
We find positive correlations between TeV and X-ray luminosities, supporting a leptonic emission scenario. 
We also find a correlation between X-ray and disk luminosities, whereas no significant correlation is found between X-ray luminosity and black hole mass.

\end{abstract}



\section{Introduction} \label{sec:intro}

Blazars are a subclass of active galactic nuclei (AGNs) characterized by relativistic jets oriented at small angles with respect to the line of sight, resulting in extreme observational properties such as strong Doppler boosting, rapid variability, and high luminosities across the electromagnetic spectrum \citep[e.g., ][]{Urry1995, Hovatta2019NewAR8701541, Raiteri2025A&ARv33}. 
Based on the equivalent width (EW) of their optical emission lines, blazars are conventionally divided into two subclasses: BL Lacertae objects (BL Lacs) and flat-spectrum radio quasars (FSRQs). 
BL Lacs exhibit weak or absent emission lines (EW $< 5~\mathrm{\mathring{A}}$), whereas FSRQs show prominent broad emission lines with EW $\geqslant 5~\mathrm{\mathring{A}}$ \citep{Stickel1991, Urry1995, Scarpa1997}.

The broadband spectral energy distributions (SEDs) of blazars are characterized by two distinct components.
The low-energy component, extending from radio to ultraviolet and in some cases X-ray bands, is widely attributed to synchrotron radiation from relativistic electrons in the jet. 
According to the synchrotron peak frequency, blazars are further classified as low-, intermediate-, and high-synchrotron-peaked sources, hereafter LSPs, ISPs, and HSPs, respectively \citep[e.g.,][]{Abdo2010ApJ716, Fan2016ApJS226, Yang2022ApJS262}.
In leptonic scenarios, the high-energy component, peaking at X-ray to $\gamma$-ray energies, is generally interpreted as inverse Compton (IC) scattering of seed photons by the same electron population.
This component can be produced through synchrotron self-Compton (SSC) emission when the seed photons are the synchrotron photons, or through external Compton (EC) emission when the seed photons originate outside the jet.
Possible external photon fields include the accretion disk \citep{Dermer1993}, the broad-line region \citep[BLR;][]{Sikora1994, Fan2006}, and the dusty torus \citep[DT;][]{Blazejowski2000, Arbeiter2002, Sokolov2005}.
Alternatively, the high-energy component can be attributed to proton synchrotron radiation and proton-induced secondary-particle cascades \citep[e.g.,][]{Mucke2001, Dimitrakoudis2012, Bottcher2013, Zheng2013, Diltz2015, Cerruti2019MNRAS483, Ouyang2025ApJ980, Xiao2026ApJ1005}.

The TeVCat~2.0 catalog\footnote{\url{http://tevcat2.uchicago.edu}} contains 303 sources associated with TeV emission, among which 84 are confirmed as blazars.
For FSRQs, TeV emission is often interpreted in terms of the EC process, with a substantial contribution from DT seed photons \citep{Xiao2024RAA24}.
For BL Lacs, the TeV emission is commonly interpreted as SSC radiation.
In such a scenario, X-ray emission takes the role of a seed photon to be scattered into the TeV band, and a consequent strong correlation between X-ray and TeV emission is expected.
In hadronic scenarios, by contrast, TeV photons can be produced through proton synchrotron radiation, neutral pion decay, or cascade emission from secondary particles \citep{Dermer2009ApJ692, Bottcher2013, Xiao2026ApJ1005}.
The TeV radiation may then depend on the proton distribution, magnetic field, target photon density, and interaction efficiency, leading to a looser correlation between X-ray and TeV emission.

TeV blazars have been widely investigated in previous works.
For instance, 
\citet{Zhang2022PASP} investigated the composition of TeV emission in blazars and separated it into the beamed and un-beamed emission;
\citet{Liang2023PASP} examined the differences between TeV and non-TeV BL Lacs in the mid-infrared band, suggesting that TeV BL Lacs are brighter and have a flatter spectrum than do the non-TeV BL Lacs in the mid-IR band. 
\citet{Ouyang2023ApJ949} explored the blazar sequence in the TeV band and suggested that the blazar sequence in the TeV band could be explained by the Doppler beaming effect. 
\citet{Wierzcholska2016MNRAS458_56} analyzed the X-ray spectral properties of a blazar sample using data from the Neil Gehrels \emph{Swift} Gamma-Ray Burst Observatory (\emph{Swift}), accounting for different neutral hydrogen absorption corrections and separating synchrotron and IC components. 
\citet{Bhatta2018A&A} investigated the hard X-ray properties of blazars observed with the \emph{Nuclear Spectroscopic Telescope Array} (\emph{NuSTAR}), focusing on timing, spectral, and cross-correlation characteristics. 
Furthermore, comprehensive X-ray catalogs of blazars have been constructed, including a soft X-ray catalog by \citet{Giommi2021MNRAS} and a hard X-ray catalog by \citet{Middei2022MNRAS}, which together provide valuable constraints on the spectral and variability behavior of these sources.

In this work, we present a systematic X-ray analysis of TeV blazars, using \emph{Swift} and \emph{NuSTAR} observations and aiming to characterize their X-ray spectral and variability properties and to explore the underlying physical processes responsible for their high-energy emission. 
The sample selection and X-ray data analysis are described in Section~\ref{Data analysis}. 
The results are presented in Section~\ref{Results}. 
We discuss the implications of our findings in Section~\ref{Discussion}, and our main conclusions are summarized in Section~\ref{Conclusion}.
Throughout this work, we adopt the $\Lambda$CDM cosmological parameters from \citet{Planck2020A&A641}, with $H_{0} = 67.7$~km~s$^{-1}$~Mpc$^{-1}$ and $\Omega_{\rm m} = 0.310$.

\section{Data analysis}\label{Data analysis}

\subsection{Sample description}

Our sample is drawn primarily from the TeVCat catalog, and the source classifications are adopted from the \emph{Fermi}-LAT 14-year Source Catalog \citep[4FGL-DR4;][]{Abdollahi2022ApJS}. 
It consists of 69 BL Lacs, 10 FSRQs, and five blazar candidates of uncertain type (BCUs). 
In addition, we include one BL Lac object, RGB J2042.1+2426 (4FGL J2042.1+2427), which is not listed in TeVCat but has been detected at TeV energies by the MAGIC telescope \citep{Acciari2020ApJS247}. 
In total, our sample comprises 85 TeV-detected blazars, including 70 BL Lacs, 10 FSRQs, and five BCUs, which are listed in Table~\ref{tab:sample}. 

All X-ray data for the sources in our sample, available up to May 2024, were retrieved from the public archive of the High Energy Astrophysics Archive Research Center (HEASARC)\footnote{\url{https://heasarc.gsfc.nasa.gov/docs/archive.html}}.

\subsection{Observation and data reduction}

All data reduction and analysis were performed using the {\ttfamily HEASoft} software package (v6.36)
provided by HEASARC. 
The procedures adopted for each X-ray instrument are described below.

\subsubsection{Swift observations}

The Neil Gehrels \emph{Swift} Observatory \citep[\emph{Swift};][]{Gehrels2004apj} was launched in 2004 and carries three instruments: the Ultraviolet and Optical Telescope \citep[UVOT;][]{Roming2005ssr}, the X-Ray Telescope \citep[XRT;][]{Burrows2004}, and the Burst Alert Telescope \citep[BAT;][]{Barthelmy2005ssr}. 

The \emph{Swift}-XRT observations were obtained in Photon Counting (PC) and Windowed Timing (WT) modes. 
XRT spectra were generated using the Swift-XRT data products provided by the UK Swift Science Data Centre at the University of Leicester \citep{Evans2009mnras}. 
Events with grades 0--12 for PC mode and 0--2 for WT mode were selected. 
Pile-up effects become significant when the count rate exceeds 0.5 counts s$^{-1}$ in PC mode or 100 counts s$^{-1}$ in WT mode, and pile-up corrections were applied following the standard procedures described by the UK Swift Science Data Centre. 
The spectra were grouped using the {\ttfamily grppha} tool to ensure at least one count per bin, allowing the use of Cash statistics \citep{Cash1979}. 
Only observations with exposure times greater than 0.3 ks and data sets with total counts exceeding 60 were included in the analysis. 
In addition, WT-mode observations with count rates greater than 0.4 counts s$^{-1}$ were retained.
Before applying the selection criteria, the data contained 12381 \emph{Swift}-XRT spectra, including 6491 PC-mode and 5890 WT-mode spectra. 
After filtering, 9833 spectra were retained for further analysis, of which 5694 were obtained in PC mode and 4139 in WT mode.

\subsubsection{NuSTAR observations}

The Nuclear Spectroscopic Telescope Array \citep[\emph{NuSTAR};][]{Harrison2013apj}, launched in 2012, operates in the hard X-ray band from 3 to 79 keV and consists of two co-aligned Focal Plane Detector Modules (FPMA and FPMB). 
The raw data were processed using the \emph{NuSTAR} Data Analysis Software (NuSTARDAS, v2.1.2) with calibration files from CALDB (version 20231017). 
Cleaned event files were produced using the {\ttfamily nupipeline} task. 
Source spectra were extracted from circular regions centered on the source, while background spectra were extracted from surrounding annular regions, with extraction radii chosen depending on the source brightness. 
We restricted the spectral analysis to the 3--60~keV energy range, where the instrumental response is optimal, and background contamination is minimized. 
The spectra were generated using the {\ttfamily nuproducts} task and grouped to at least one count per bin to allow for Cash fitting. 
Only observations with exposure times greater than 1.0 ks were included.
A total of 146 \emph{NuSTAR} spectra were initially available. 
After selection, 140 spectra were retained for further analysis.

\subsection{Spectral fitting}

Spectral fitting was performed using the {\ttfamily XSPEC} package \citep{Arnaud1996ASPC101}. 
Solar abundances were set to {\ttfamily wilm} \citep{Wilms2000}, and photoelectric absorption cross-sections were adopted from {\ttfamily vern} \citep{Verner1996}. 
Each spectrum was fitted using two models: a power-law and a log-parabolic model, both multiplied by the Galactic absorption model {\ttfamily tbabs} \citep{Wilms2000}.  
The Galactic total hydrogen column density was fixed to the value reported by \citet{HI4PI2016}, as shown in Table~\ref{tab:sample}. 

The power-law model is defined as
\begin{equation}
\frac{\mathrm{d}N}{\mathrm{d}E} = K_{0} E^{-\Gamma},
\end{equation}
where $K_{0}$ is the normalization and $\Gamma$ is the photon index. 
The log-parabolic model is given by
\begin{equation}
\frac{\mathrm{d}N}{\mathrm{d}E} = K_{0} \left( \frac{E}{E_{0}} \right)^{-\left[ \alpha + \beta \log \left( \frac{E}{E_{0}} \right) \right]},
\end{equation}
where $E_{0}$ is the pivot energy, fixed at 1~keV for \emph{Swift}-XRT 
and at 5~keV for \emph{NuSTAR} spectra, $\alpha$ is the photon index at $E_{0}$, and $\beta$ is the curvature parameter. 
For \emph{NuSTAR} spectra, a cross-calibration constant was included for FPMB relative to FPMA, yielding values within a few percent in all cases, consistent with previous calibration studies \citep{Madsen2015ApJS}. 
Parameter uncertainties were estimated at the 68.3\% confidence level, corresponding to $1\sigma$ for one parameter of interest.

The power-law model is nested within the log-parabolic model, with the latter including one additional free parameter, the curvature parameter $\beta$. 
We used the F-test to assess whether the improvement in the fit obtained by introducing the curvature parameter was statistically significant\citep[e.g.,][]{Bevington2003drea_book_F-Test}. 
We adopted the log-parabolic model only when the F-test probability was below $0.01$, corresponding to a 99\% confidence level.
Otherwise, the power-law model was selected.

Finally, for each source we derived the unabsorbed integrated fluxes in the 0.3--10~keV band for \emph{Swift}-XRT and in the 3--60~keV band for \emph{NuSTAR}, along with the corresponding best-fitting spectral parameters.
For each source, we compiled a complete X-ray data table containing all \emph{Swift}-XRT and \emph{NuSTAR} observations used in this work, which is shown in Appendix~\ref{Appendix:Xray_Data}.

\section{Results}\label{Results}

\subsection{Long-term light curves}

\begin{figure}
    \centering
    \includegraphics[width=3.2 in]{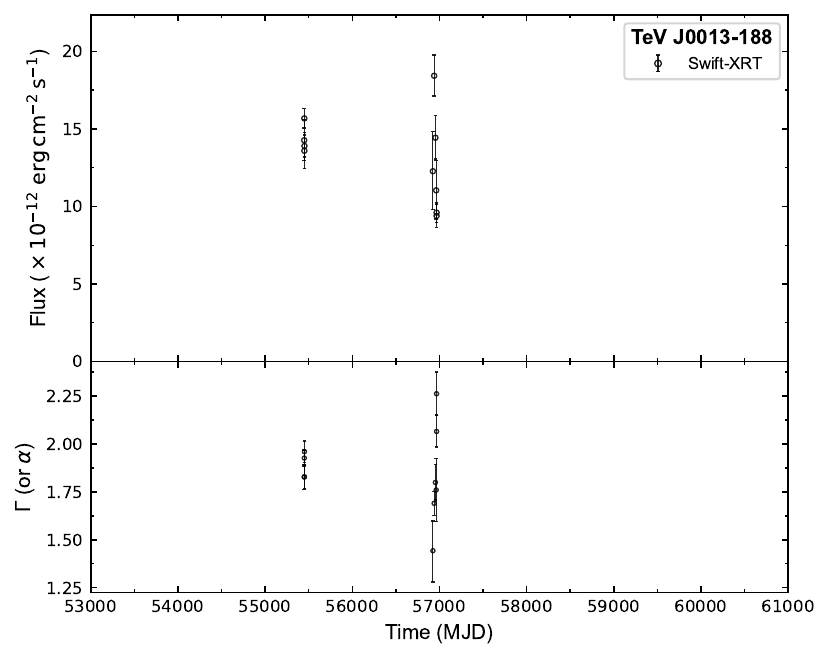}
    \includegraphics[width=3.2 in]{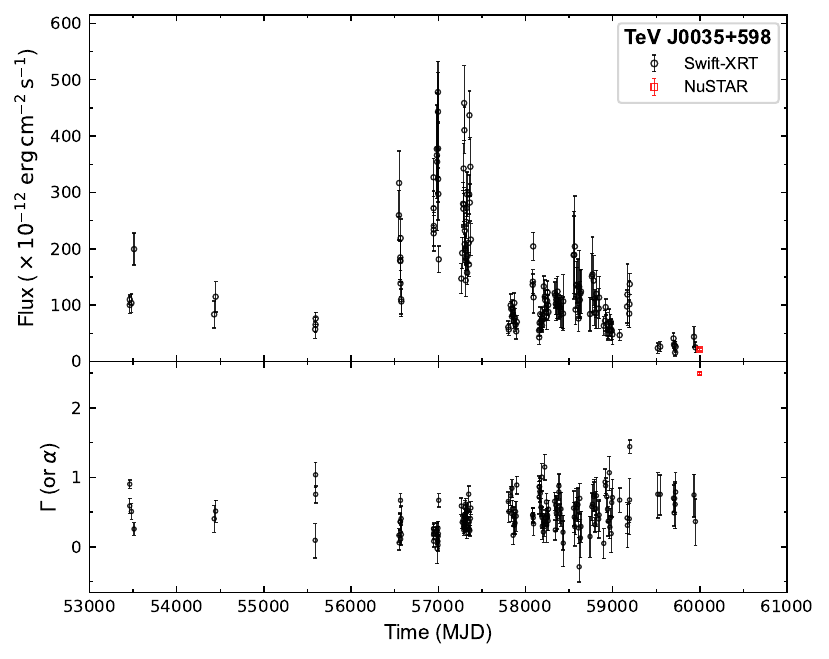}
    \includegraphics[width=3.2 in]{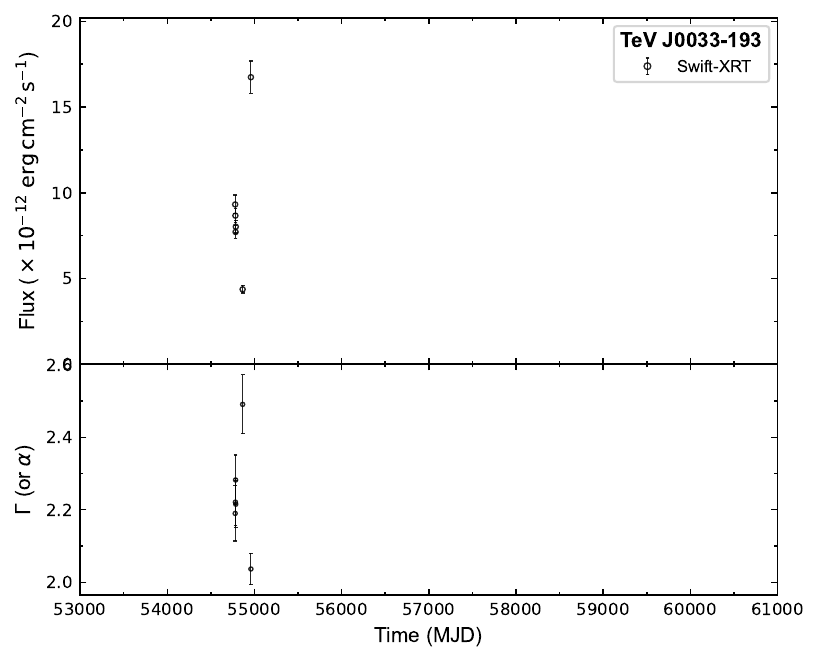}
    \includegraphics[width=3.2 in]{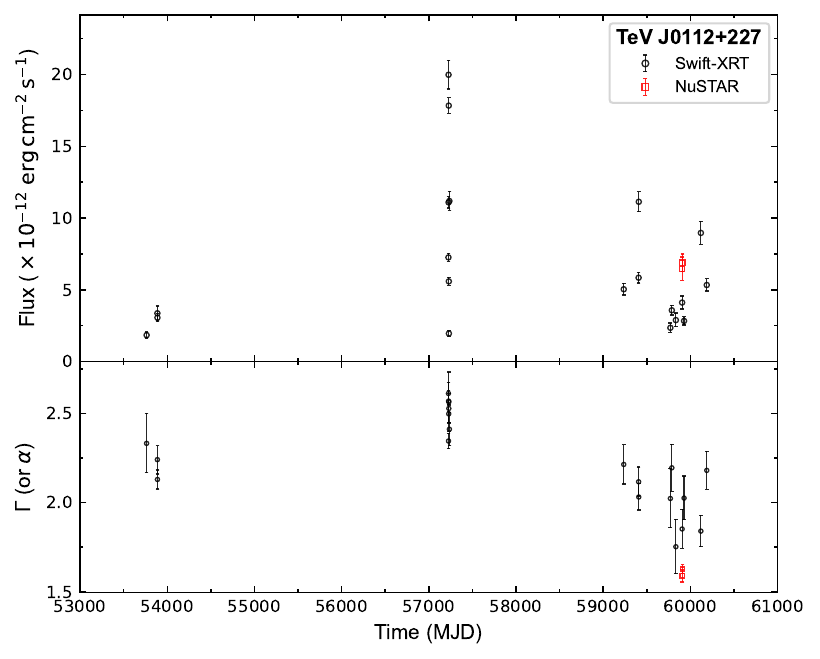}
    \caption{Long-term X-ray light curves of TeV blazars. 
    The black open circles denote \emph{Swift}-XRT observations
    and the red open squares denote \emph{NuSTAR} observations. 
    For each source, the upper panel shows the unabsorbed flux and the lower panel shows the photon index. \\
    (Only four representative sources are shown as examples.)}
    \label{fig:lightcurve}
\end{figure}

For each TeV blazar in our sample, we derived the unabsorbed X-ray fluxes integrated over 0.3--10~keV for \emph{Swift}-XRT and over 3--60~keV for \emph{NuSTAR} observations, together with the corresponding photon indices from the best-fitting spectral models. 
The resulting long-term light curves are shown in Figure~\ref{fig:lightcurve}. 
For convenience, we combine the \emph{Swift}-XRT measurements (PC and WT modes) and refer to them as \emph{soft X-ray} observations, while \emph{NuSTAR} measurements are referred to as \emph{hard X-ray} observations. 
For each source, we computed the arithmetic mean flux and photon index separately for the soft and hard X-ray observations; these values are listed in Table~\ref{tab:sample}.

\subsection{Spectral energy distributions}

\begin{figure}
    \centering
    \includegraphics[width=3.1 in]{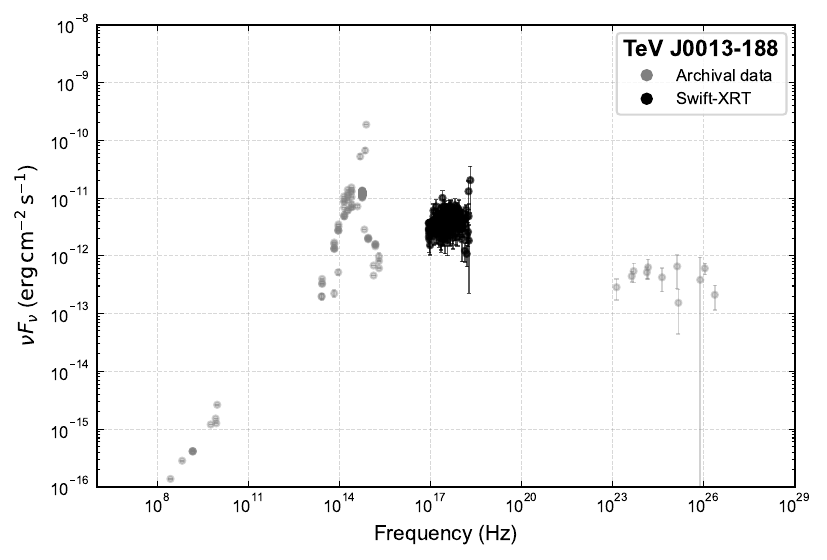}
    \includegraphics[width=3.1 in]{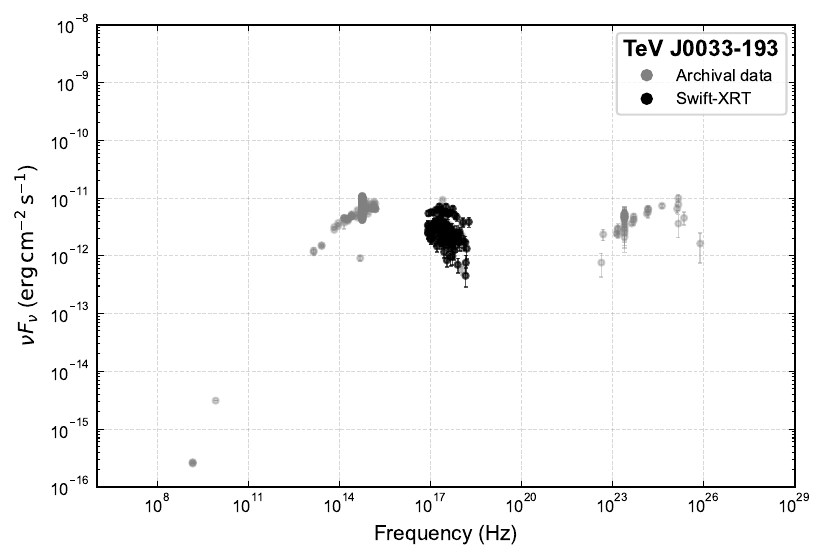}
    \includegraphics[width=3.1 in]{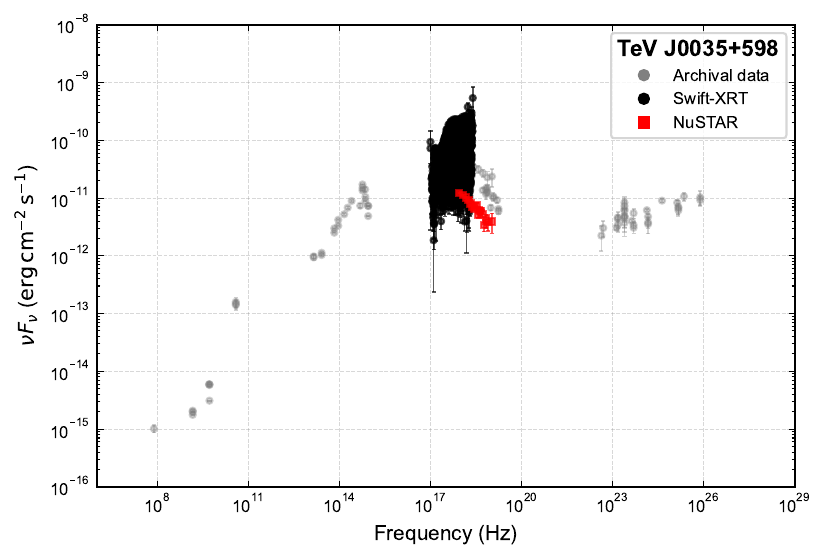}
    \includegraphics[width=3.1 in]{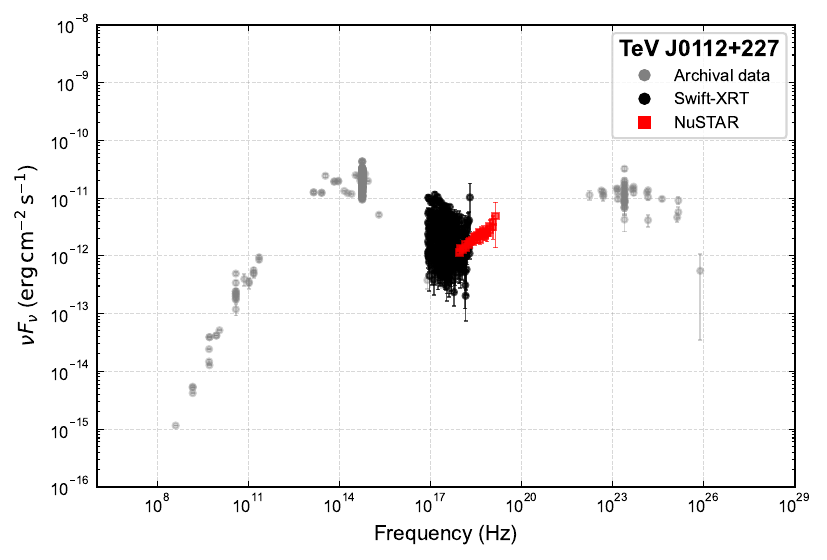}
    \caption{Spectral energy distributions of TeV blazars. 
    The black-filled circles denote \emph{Swift}-XRT spectra, the red-filled squares denote \emph{NuSTAR} (FPMA) spectra, and the gray-filled circles denote archival data compiled by SSDC. \\
    (Only four representative sources are shown as examples.)}
    \label{fig:SED}
\end{figure}

Using the best-fitting spectral models obtained in Section \ref{Data analysis}, we constructed X-ray spectral energy distributions (SEDs) for each TeV blazar to examine spectral variability and evolution. 
The X-ray SEDs are shown in Figure~\ref{fig:SED}. 
The spectra are corrected for Galactic absorption in a manner consistent with our spectral fitting using the cross-sections from \citet{Morrison1983ApJ} and the Galactic hydrogen column densities listed in Table~\ref{tab:sample}. 
For comparison, archival broadband SED data were retrieved from the Space Science Data Center (SSDC)\footnote{\url{https://www.ssdc.asi.it/}}. 
For clarity, only FPMA spectra are shown for \emph{NuSTAR} observations.

\subsection{Flux distribution of TeV blazars}\label{sec:fluxdist}

\begin{figure}
    \centering
    \includegraphics[width=3.1 in]{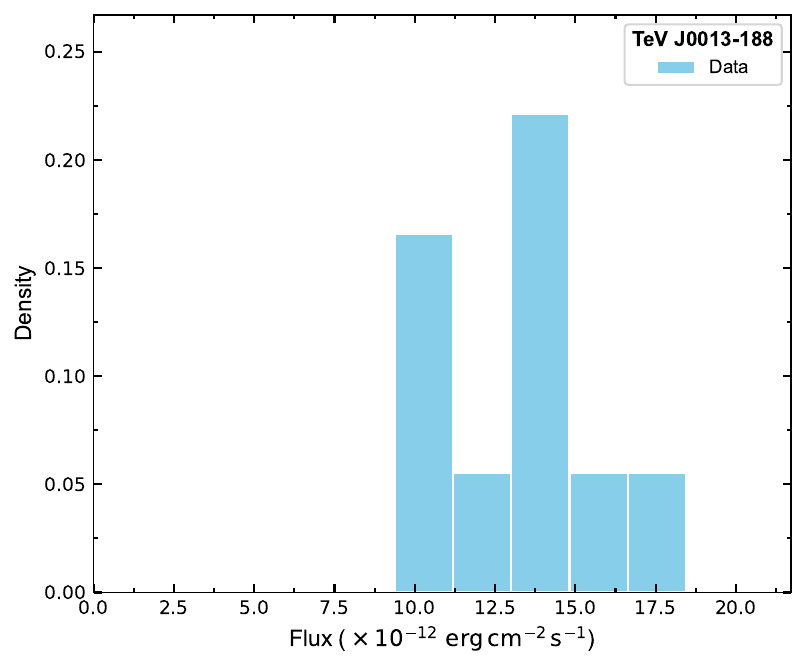}
    \includegraphics[width=3.1 in]{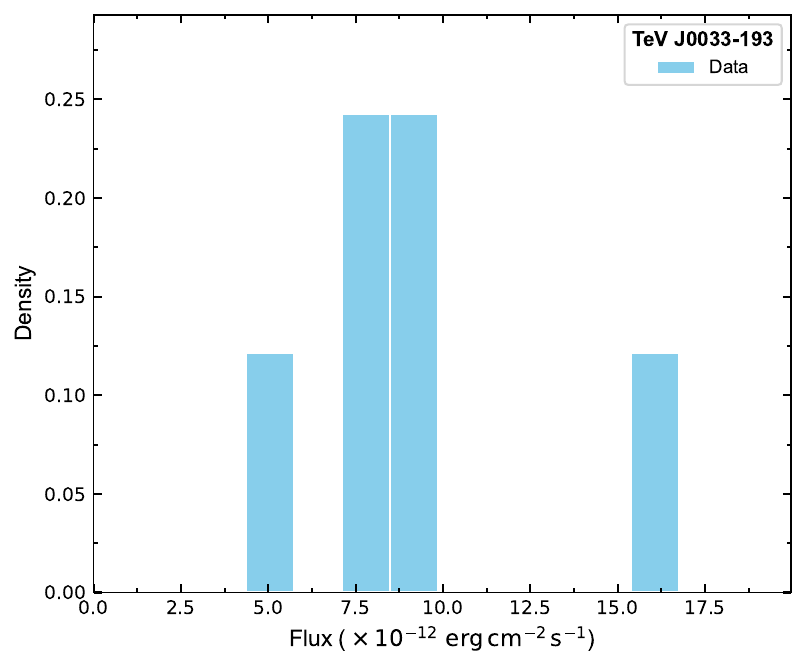}
    \includegraphics[width=3.1 in]{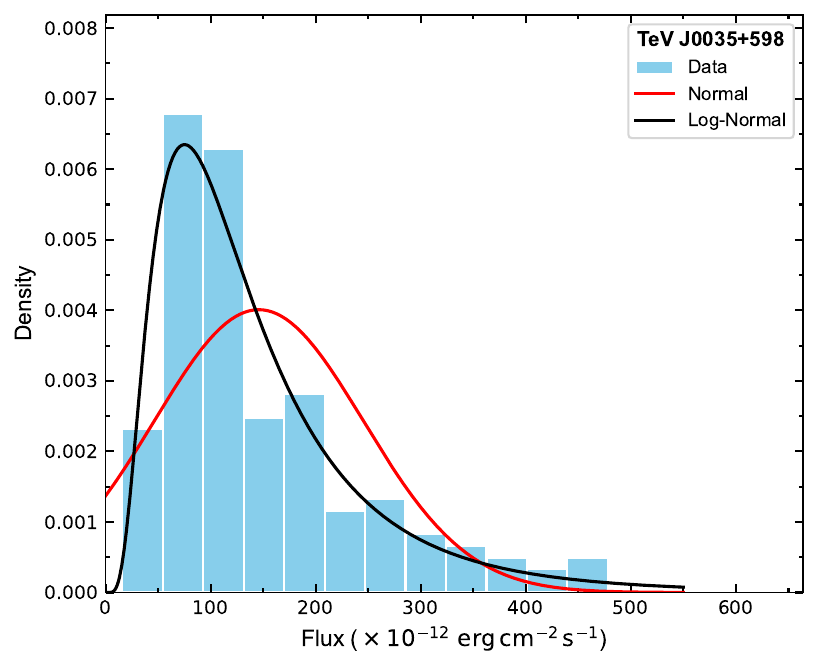}
    \includegraphics[width=3.1 in]{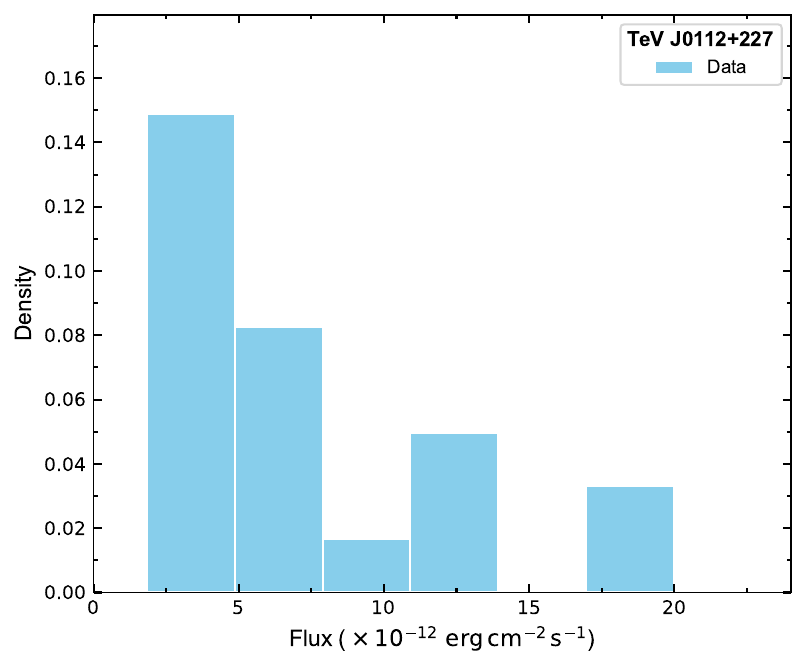}
    \caption{Soft X-ray flux distributions of TeV blazars. 
    Histograms are shown for all sources displayed, while normal and log-normal probability density functions are fitted only for sources with sufficiently well-sampled soft X-ray light curves. 
    The red and black solid curves denote the best-fitting normal and log-normal PDFs, respectively. \\
    (Only four representative sources are shown as examples.)}
    \label{fig:PDF}
\end{figure}

The X-ray flux distribution provides a useful statistical probe of the variability processes in blazars. 
However, the number of available X-ray observations varies significantly across the sources in our sample, and many TeV blazars are only sparsely observed. 
To ensure meaningful statistical characterization of flux distributions, we restrict the following analysis to sources with sufficiently well-sampled soft X-ray light curves, while still describing variability properties for the full sample elsewhere in this work.
Specifically, we consider sources with more than 30 soft X-ray observations to have adequate sampling for constructing flux histograms and fitting probability density functions (PDFs). 
For these well-sampled sources, we constructed histograms of the soft X-ray flux measurements and fitted them with both normal and log-normal PDFs, as shown in Figure~\ref{fig:PDF}. 
We compared the goodness of fit using the reduced $\chi^{2}$ values to identify the preferred statistical description. 
We find that, for the majority of these sources, the soft X-ray flux distributions are better described by a log-normal PDF, while TeV~J0232+202 is better described by a normal distribution. 
The best-fitting parameters for both models are summarized in Table~\ref{tab:PDF_fitting}.

We do not perform an analogous analysis for hard X-ray flux distributions because the number of hard X-ray observations is generally much smaller than that of soft X-ray observations across the sample. 
Even for the most extensively observed source in hard X-rays (TeV~J1104+382), the available number of measurements ($\sim$30) remains insufficient for robust distribution fitting. 
As a result, the flux distribution analysis presented here focuses exclusively on the soft X-ray band.

\subsection{Flux--index correlations of TeV blazars}

Correlations between X-ray spectral shape and flux provide a valuable diagnostic of spectral evolution and can offer insights into the underlying radiation processes. 
We examined the relationships between X-ray fluxes and the corresponding photon indices for the TeV blazars in our sample. 
Following the same sampling considerations as in Section~\ref{sec:fluxdist}, we perform the soft X-ray flux--index correlation analysis only for sources with sufficiently well-sampled light curves.
For the hard X-ray band, we require more than 10 \emph{NuSTAR} observations before performing the correlation analysis. 

For each eligible source, we quantified the strength of the linear relationship between flux and photon index using the Pearson correlation test and performed ordinary least-squares (OLS) regression following \citet{Isobe1990ApJ}. 
We treat correlations as statistically significant when the Pearson test yields a probability of $p < 0.05$, under the null hypothesis of no correlation. 
The best-fitting regression parameters are summarized in Table~\ref{tab:linear_regression}. 

\begin{figure}
    \centering
    \includegraphics[width=3.1 in]{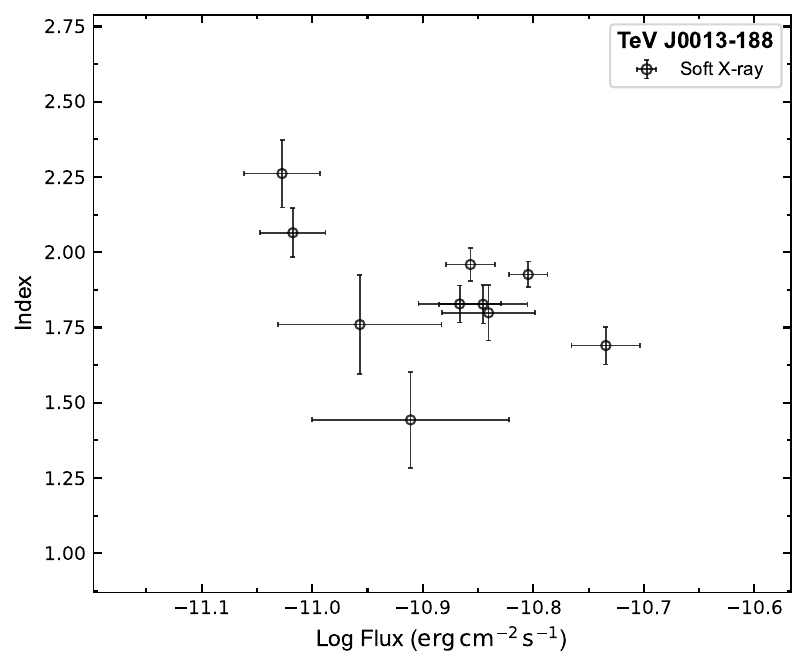}
    \includegraphics[width=3.1 in]{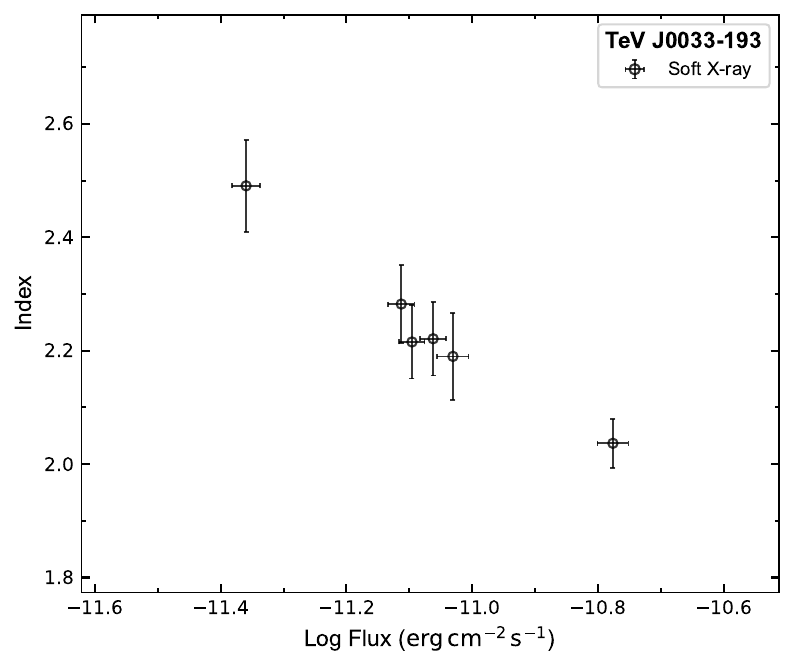}
    \includegraphics[width=3.1 in]{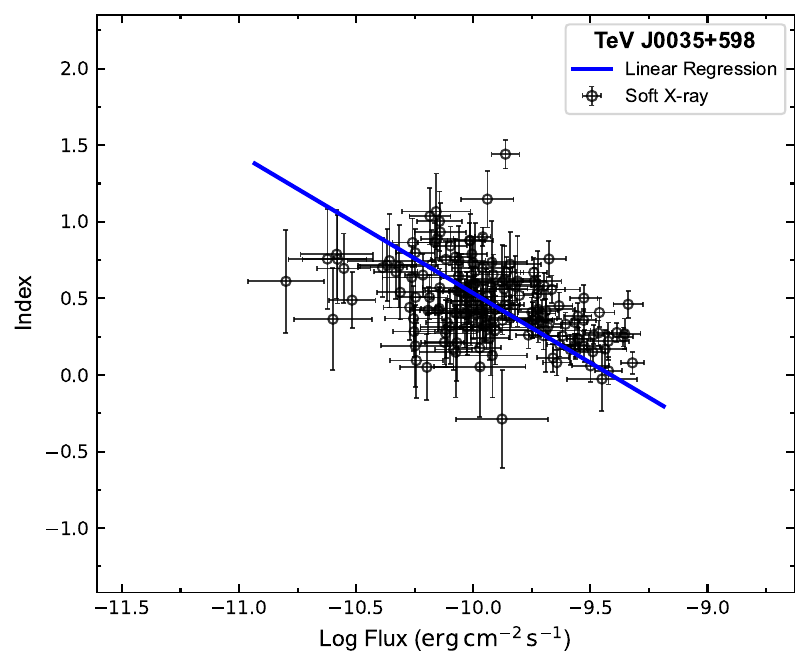}
    \includegraphics[width=3.1 in]{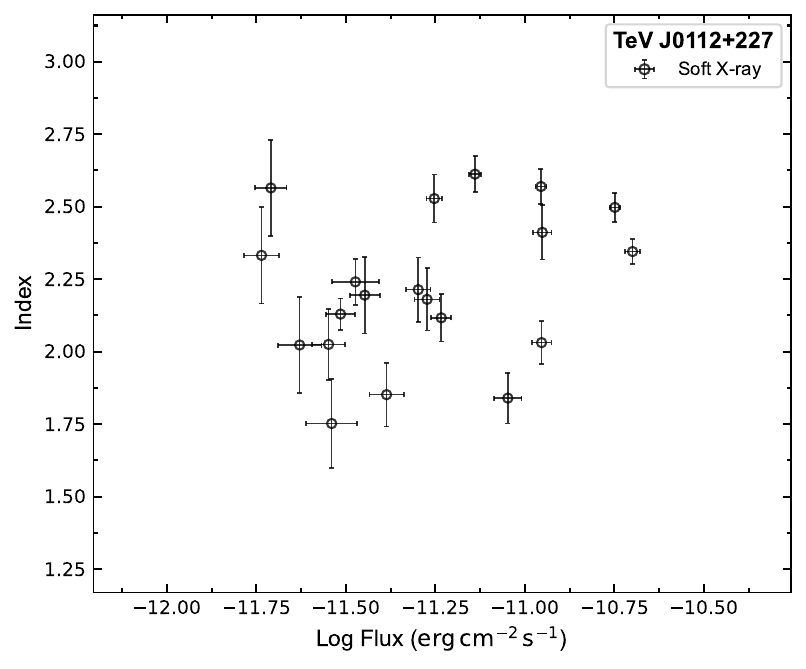}
    \caption{Soft X-ray flux versus photon index for TeV blazars. 
    Data points are shown for all sources displayed, while ordinary least-squares (OLS) linear regression is performed only for sources with sufficiently well-sampled soft X-ray light curves. 
    The blue solid line denotes the resulting regression. \\
    (Only four representative sources are shown as examples.)}
    \label{fig:Flux_vs_Index_Soft}
\end{figure}

\begin{figure}
    \centering
    \includegraphics[width=3.1 in]{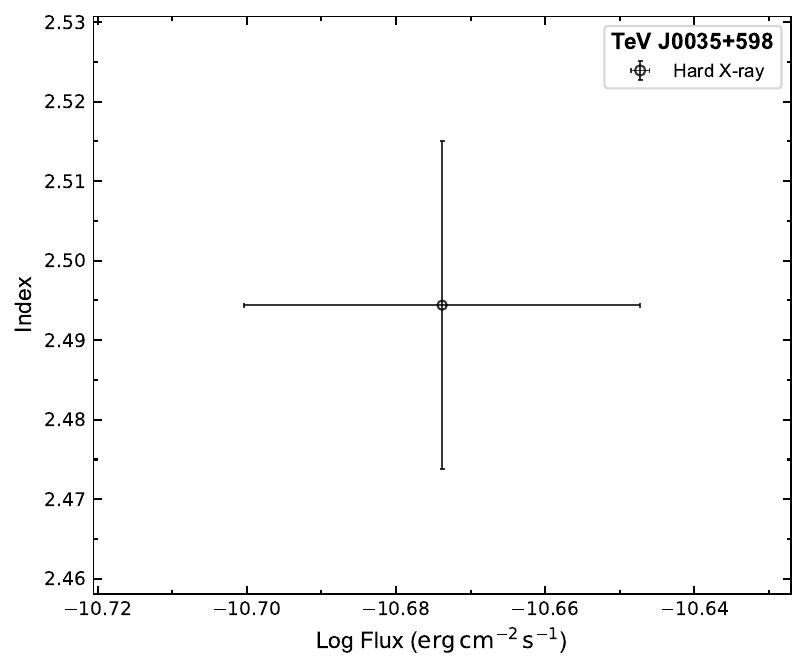}
    \includegraphics[width=3.1 in]{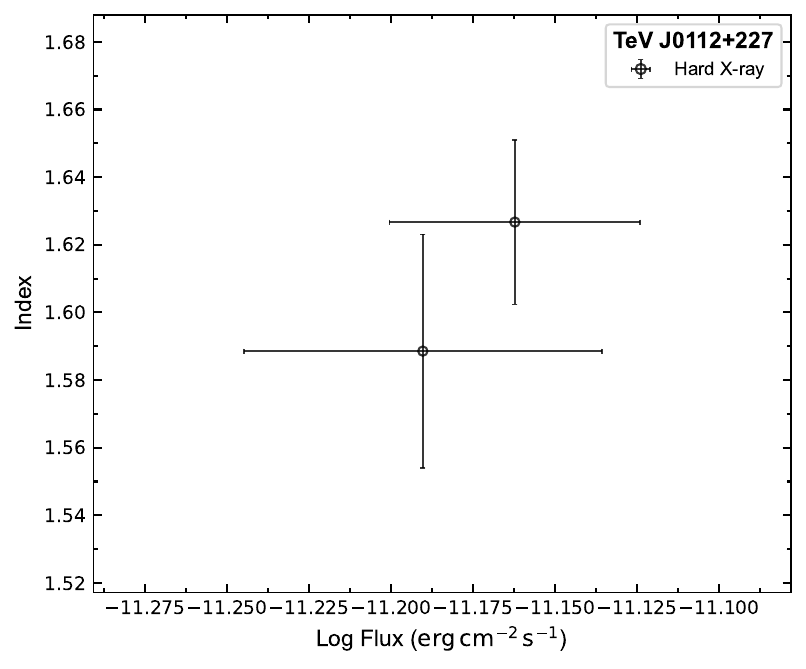}
    \includegraphics[width=3.1 in]{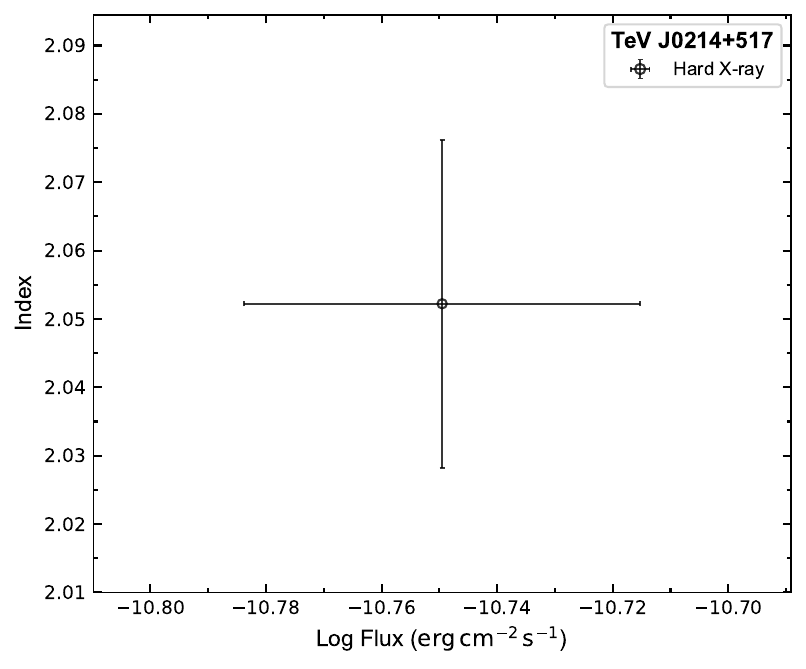}
    \includegraphics[width=3.1 in]{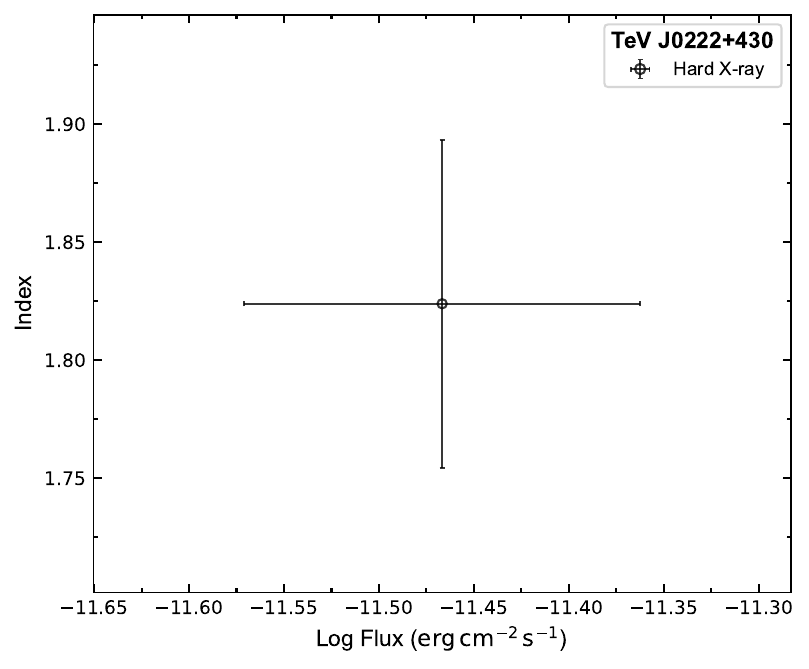}
    \caption{Hard X-ray flux versus photon index for TeV blazars. 
    Data points are shown for all sources displayed, while ordinary least-squares (OLS) linear regression is performed only for sources with sufficiently well-sampled hard X-ray light curves. 
    The blue solid line denotes the resulting regression. \\
    (Only four representative sources are shown as examples.)}
    \label{fig:Flux_vs_Index_Hard}
\end{figure}

The flux--index relations derived from soft X-ray measurements are shown in Figure~\ref{fig:Flux_vs_Index_Soft}, while those based on hard X-ray measurements are shown in Figure~\ref{fig:Flux_vs_Index_Hard}. 
Sources with fewer observations are not included in the regression analysis, as limited sampling can obscure intrinsic correlations and lead to unreliable slope estimates.

\section{Discussion}\label{Discussion}

\subsection{Correlation between X-ray luminosity and TeV luminosity}

The correlation between X-ray and TeV luminosities could probe whether the two bands are produced by physically connected emission processes.
We therefore compare the X-ray luminosities derived in this work with the TeV luminosities compiled from \citet{Ouyang2023ApJ949}.
The X-ray fluxes are obtained from the spectral fitting described in Section~\ref{Data analysis}.
The X-ray luminosity is then calculated as
\begin{equation}
L_{\rm X} = 4\pi d_{\rm L}^{2} F_{\rm X} (1+z)^{\Gamma_{\rm eff}-2},
\label{Equ:L_X}
\end{equation}
where $F_{\rm X}$ is the unabsorbed integrated X-ray flux, $d_{\rm L}$ is the luminosity distance derived from the source redshift, $z$ is the redshift, and $(1+z)^{\Gamma_{\rm eff}-2}$ is the $K$-correction.
For spectra fitted with a power-law model, we adopt $\Gamma_{\rm eff}=\Gamma$, while for spectra fitted with a log-parabolic model, we adopt $\Gamma_{\rm eff}=\alpha$, where $\alpha$ is the photon index at the reference energy.
For each source, the X-ray luminosity was calculated for each individual observation using Equation~\ref{Equ:L_X}.
We then computed the arithmetic mean of these luminosities and adopted it as the representative X-ray luminosity of the source.
The mean soft X-ray luminosity was derived from all available \emph{Swift}-XRT observations in the 0.3--10~keV band, while the mean hard X-ray luminosity was derived from all available \emph{NuSTAR} observations in the 3--60~keV band.
These mean X-ray luminosities were then used to investigate their correlations with the TeV luminosities.

\begin{figure}
    \centering
    \includegraphics[width=7 in]{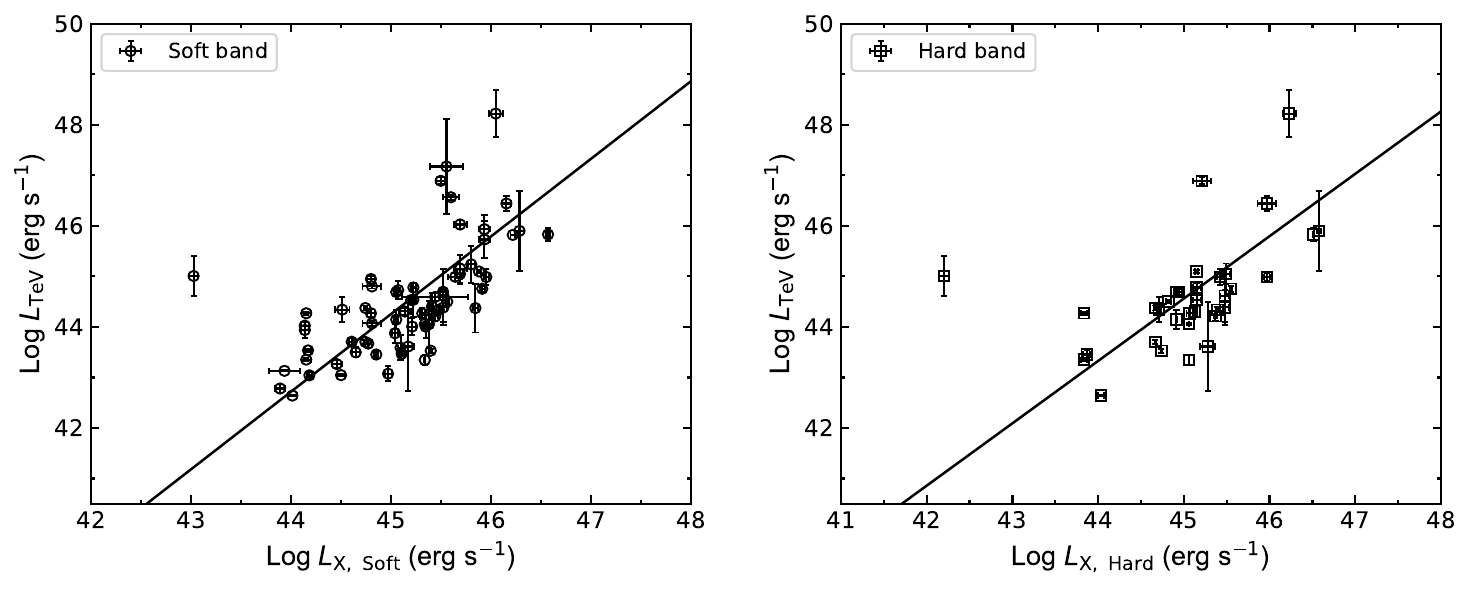}
    \caption{The TeV luminosity as a function of the soft (left) and hard (right) X-ray luminosities. 
    The black line is the regression line.}
    \label{fig:XrayLum_vs_TeVLum}
\end{figure}

Figure~\ref{fig:XrayLum_vs_TeVLum} shows the TeV luminosity as a function of the soft and hard X-ray luminosities.
We use OLS regression and the Pearson test to quantify the correlations; the results are listed in Table~\ref{tab:OLS_Pearson}.
We find significant positive correlations between TeV luminosity and both soft and hard X-ray luminosities.
This result supports a physical connection between the X-ray and TeV emissions in the sample.
In the leptonic framework, such a connection is expected, as the X-ray and TeV emissions are produced by the same population of relativistic electrons.
For many TeV blazars, especially HSPs, the X-ray is usually dominated by synchrotron radiation from high-energy electrons, while the TeV emission is mainly produced through inverse Compton scattering by the same electron population.
For LSPs and some ISPs, however, the X-ray may instead originate from the low-energy part of the inverse Compton component or the transition region between the synchrotron and inverse Compton components \citep[e.g.,][]{Finke2008ApJ686, Dermer2009ApJ692}.
Even in these cases, a positive luminosity correlation can arise if both bands respond to the changes in physical conditions in the emitting region, such as particle density or Doppler boosting.

By contrast, in hadronic scenarios, the TeV emission can originate from several processes, such as proton synchrotron radiation, neutral pion decay, and cascade emission from secondary particles.
In these cases, the TeV luminosity is controlled not only by the electron population and leptonic radiative component responsible for the X-ray emission, but also by the proton distribution, magnetic field strength, target photon density, and efficiency of photohadronic interactions \citep[e.g.,][]{Bottcher2013, Cerruti2019MNRAS483}.
As a result, a tight or direct correlation between X-ray and TeV luminosities is not necessarily expected.

\subsection{Correlation between X-ray luminosity and central engine}

The X-ray emission of blazars is mainly produced in relativistic jets, and the jet power of blazars is connected with the central engine, including the black hole and accretion system \citep{Blandford1977MNRAS179, Blandford1982MNRAS199}.
Previous studies have also suggested a close connection between accretion power and jet power in blazars \citep[e.g.,][]{Ghisellini2009MNRAS399, Ghisellini2014Natur515, Xiao2022ApJ925, Xie2024ApJ976}.
Motivated by this disk--jet connection, we further examine whether the X-ray luminosity is related to the basic properties of the central engine of TeV blazars.
Specifically, we compare the X-ray luminosity with the black hole mass, $M_{\rm BH}$, and accretion disk luminosity, $L_{\rm disk}$, whose values are collected from \citet{Xie2024ApJ976}.

\begin{figure}
    \centering
    \includegraphics[width=7 in]{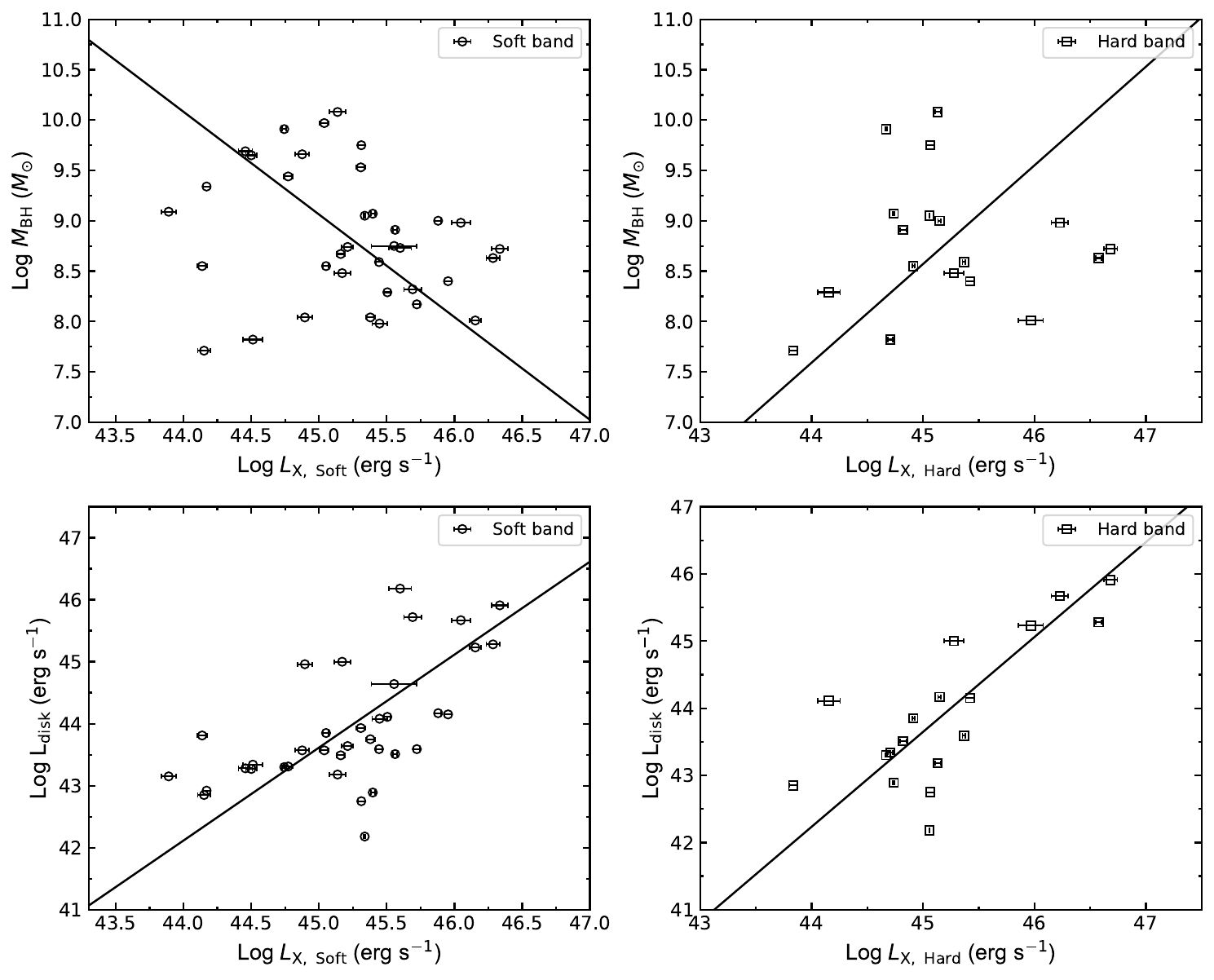}
    \caption{The black hole mass (upper) and disk luminosity (lower) as a function of the soft (left) and hard (right) X-ray luminosities. 
    The black line is the regression line.}
    \label{fig:XrayLum_vs_CentralEngine}
\end{figure}

Figure~\ref{fig:XrayLum_vs_CentralEngine} shows the black hole mass and disk luminosity as functions of the soft and hard X-ray luminosities, with the corresponding correlation results listed in Table~\ref{tab:OLS_Pearson}. 
No significant correlation is found between black hole mass and X-ray luminosity.
This suggests that the X-ray luminosity of TeV blazars is unlikely to be governed solely by black hole mass. 
This lack of dependence is reasonable because the observed X-ray emission is dominated by the relativistic jet rather than by the disk--corona system commonly seen in radio-quiet AGNs.
Relativistic beaming, viewing angle, magnetic field strength, and particle acceleration efficiency can all broaden or obscure any intrinsic relation with black hole mass.
In contrast, significant positive correlations are found between disk luminosity and both soft and hard X-ray luminosities.
These results suggest that sources with higher disk luminosities tend to have stronger X-ray emission, which may reflect an underlying disk--jet connection. 
A higher disk luminosity generally indicates a larger accretion power, which can provide more energy for jet launching and particle acceleration, producing more powerful non-thermal radiation.

\subsection{Timing analysis}

\subsubsection{Variability}

We quantified the variability in excess of the measurement uncertainties for each source using the fractional variability amplitude, \(F_{\rm var}\), which is commonly used in AGN variability studies and is defined as \citep{Vaughan2003MNRAS}
\begin{equation}
    F_{\rm var} = \sqrt{ \frac{\frac{1}{N-1} \sum^{N}_{i=1} \left( x_{i} - \overline{x} \right)^{2} - \overline{\sigma_{\rm err}^{2}}}{\mu^{2}} },
    \label{Eq:F_var}
\end{equation}
where $N$ is the number of data points, and $x_{i}$ is the $i$-th flux measurement, $\overline{x}$ and $\overline{\sigma_{\rm err}^{2}}$ are the mean flux and the mean square flux uncertainty, respectively.
The uncertainty on $F_{\rm var}$, denoted by $\sigma_{F_{\rm var}}$, was calculated as \citep{Vaughan2003MNRAS, Poutanen2008MNRAS389} 
\begin{equation}
    \sigma_{F_{\rm var}} = \sqrt{F_{\mathrm{var}}^{2}+\sqrt{\left(\sqrt{\frac{2}{N}} \cdot \frac{\overline{\sigma_{\rm err}^{2}}}{\overline{x}^{2}}\right)^{2}+\left(\sqrt{\frac{\overline{\sigma_{\rm err}^{2}}}{N}} \cdot \frac{2 F_{\mathrm{var}}}{\overline{x}}\right)^{2}}}-F_{\mathrm{var}} .
\end{equation}

The results are listed in Table~\ref{tab:sample}. 
When the numerator in Equation~\eqref{Eq:F_var} is negative, no positive fractional variability is obtained, indicating that the observed variability is weak relative to the measurement uncertainties.

We classify a source as variable when $F_{\rm var} \geqslant 3\sigma_{F_{\rm var}}$. 
With this criterion, we identify 78 sources as variable in soft X-rays and 14 sources as variable in hard X-rays. 
This confirms that X-ray variability is common in TeV blazars.
However, we caution that X-ray monitoring is often observationally biased toward bright and/or flaring states. 
In addition, many X-ray light curves are irregularly sampled and contain substantial observational gaps, which can affect variability diagnostics and may cause some sources to appear less variable than they actually are.

\subsubsection{Power spectral density}

The long-term X-ray light curves of TeV blazars exhibit stochastic variability spanning a wide range of amplitudes, with some sources varying by orders of magnitude.  
To characterize variability as a function of temporal frequency, we constrain the underlying power spectral density (PSD) properties.
Because reliable PSD estimation requires a sufficient number of measurements, we restrict this analysis to sources with well-sampled soft X-ray light curves, as in the same procedure in Section~\ref{sec:fluxdist}.

The sparse and uneven sampling of the long-term X-ray light curves in our sample, as well as the effects of aliasing and red-noise leakage, can distort the intrinsic PSDs.
We therefore characterized the PSD slopes using the multiple fragments variance function \citep[MFVF;][]{Kastendieck2011A&A531}, combined with a forward-modeling procedure analogous to the power spectral response (PSRESP) method \citep{Uttley2002MNRAS}.
The MFVF characterizes the variability of the source at different timescales by recursively dividing the light curve into two equal segments and computing the variance of each segment until a predefined minimum timescale is reached.
We assumed an intrinsic power-law PSD, $P(f)\propto f^{-\beta_{\rm PSD}}$, and tested PSD slopes over the range $\beta_{\rm PSD}=0.0-3.0$ in steps of 0.1. 
For each trial value of $\beta_{\rm PSD}$, we generated 1000 artificial light curves using the method of \citet{Emmanoulopoulos2013MNRAS}, which reproduces both the assumed PSD and the best-fitting flux PDF derived in Section~\ref{sec:fluxdist}. 
To account for red-noise leakage, each artificial light curve was initially generated with a duration 100 times longer than that of the observed light curve.
Each simulated light curve was then resampled to match the temporal sampling of the corresponding observed light curve.
For each simulated light curve, we calculated the MFVF in the same way as for the observed data.
Thus, for a given trial slope $\beta_{\rm PSD}$ and MFVF bin $i$, the 1000 simulations yielded a distribution of simulated MFVF values.
For each bin, we compared the observed MFVF value with the corresponding simulated distribution by evaluating the probability density of this distribution at the observed MFVF value, denoted as $p(\beta_{\rm PSD}, f_i)$.
A larger value of $p(\beta_{\rm PSD}, f_i)$ indicates that the observed MFVF value is more consistent with the MFVF values produced by the simulations for the trial PSD slope.
The likelihood of each trial PSD slope was then evaluated by combining the probability densities over all MFVF bins, with the log-likelihood defined as
\begin{equation}
    \mathcal{L}(\beta_{\rm PSD}) = \sum_{i=1} \ln p(\beta_{\rm PSD}, f_i).
    \label{eq:log-likelihood_PSD}
\end{equation}
The value of $\beta_{\rm PSD}$ that maximizes $\mathcal{L}(\beta_{\rm PSD})$ was adopted as the best-fitting PSD slope.

\begin{figure}
    \centering
    \includegraphics[width=3.1 in]{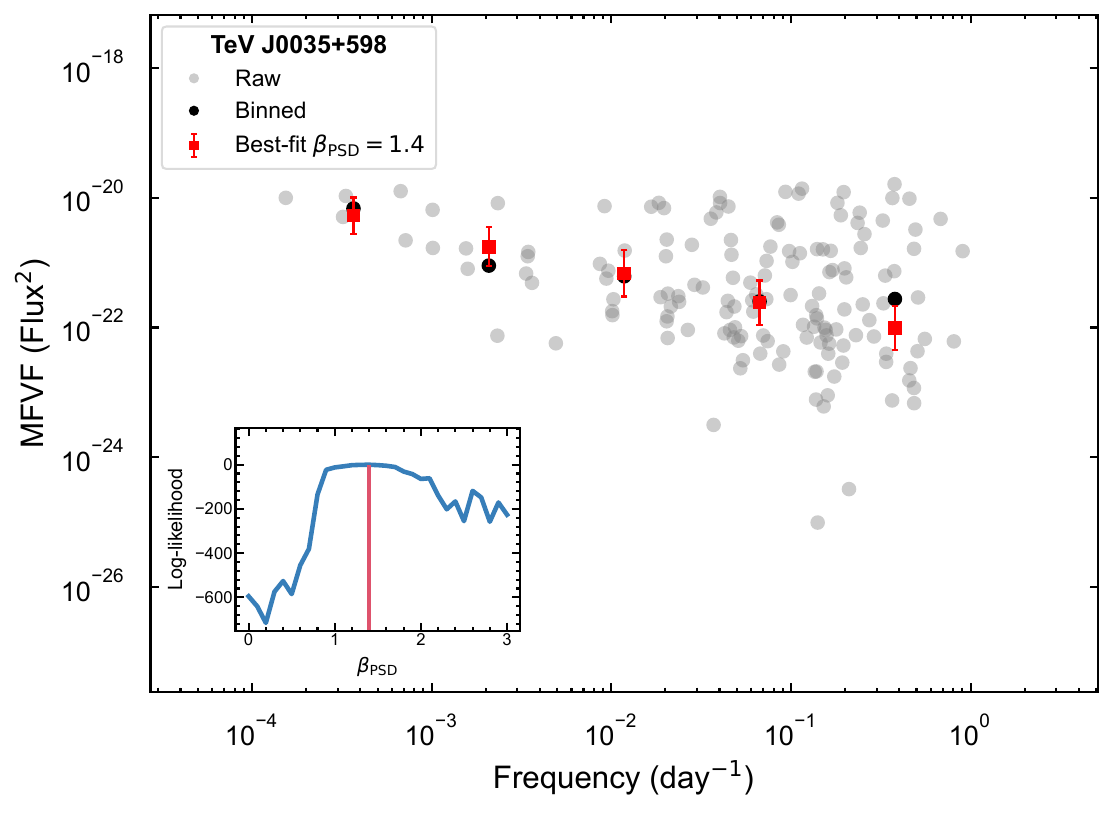}
    \includegraphics[width=3.1 in]{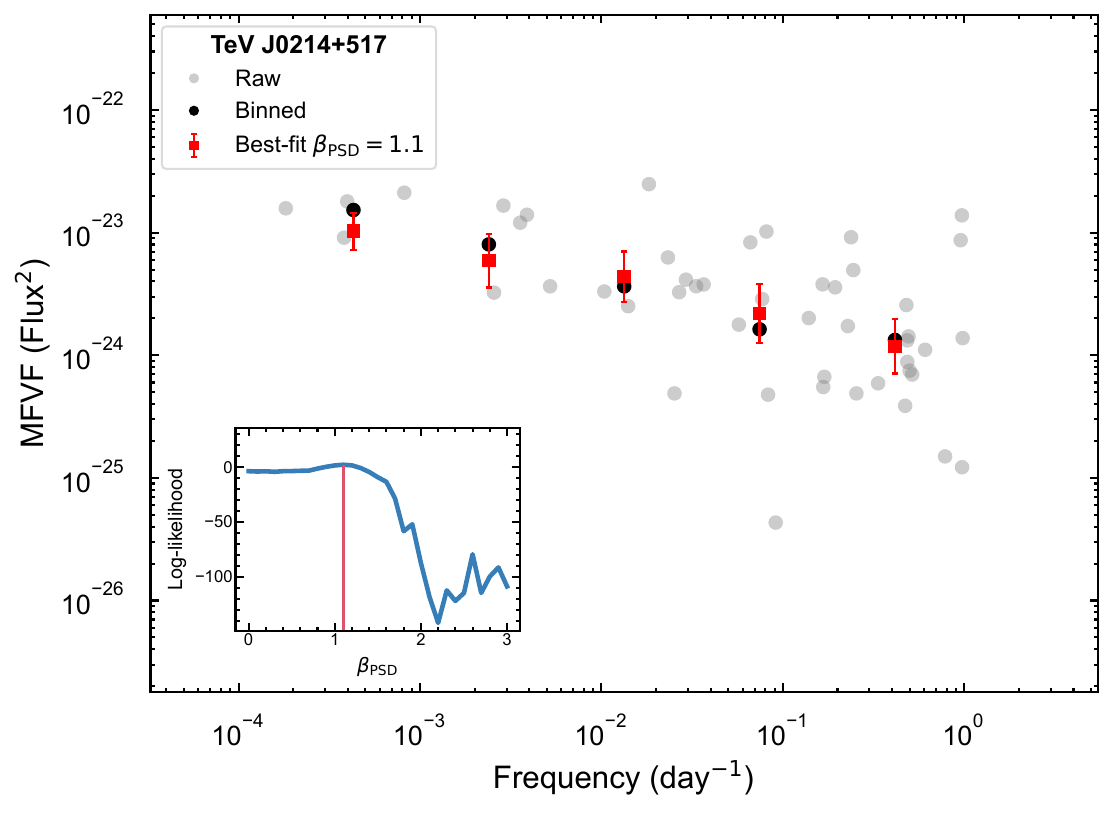}
    \includegraphics[width=3.1 in]{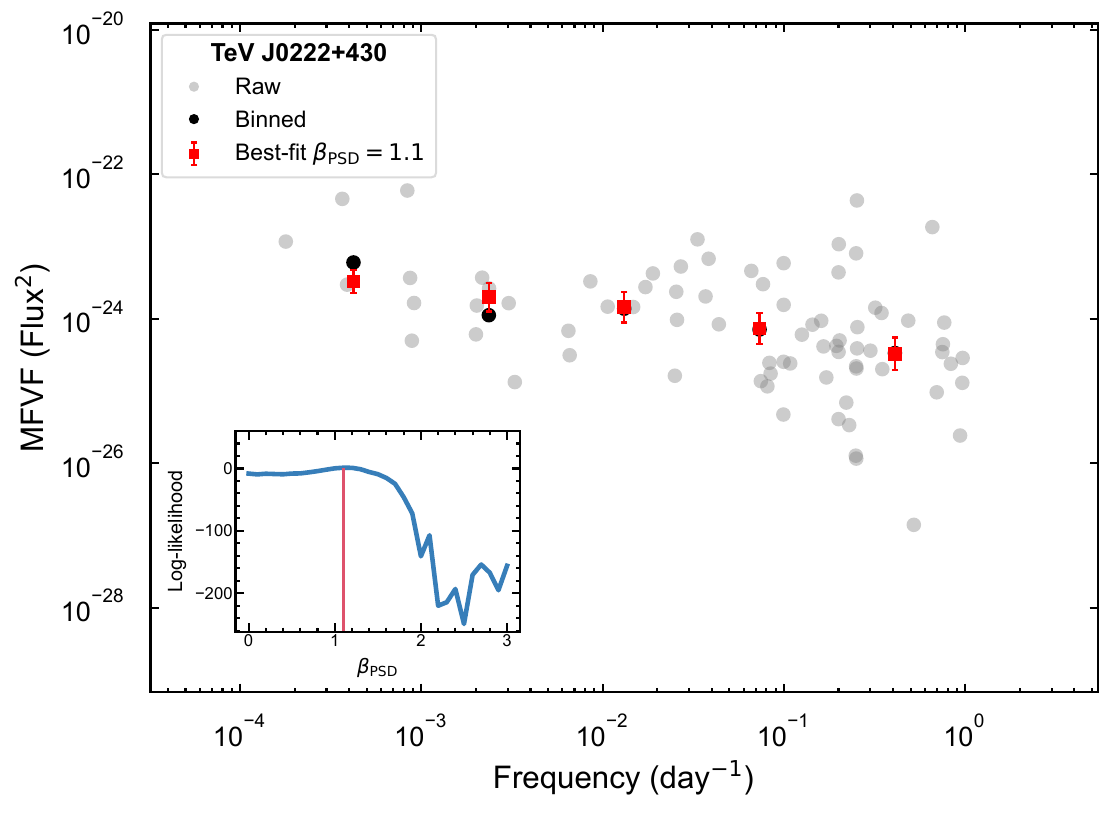}
    \includegraphics[width=3.1 in]{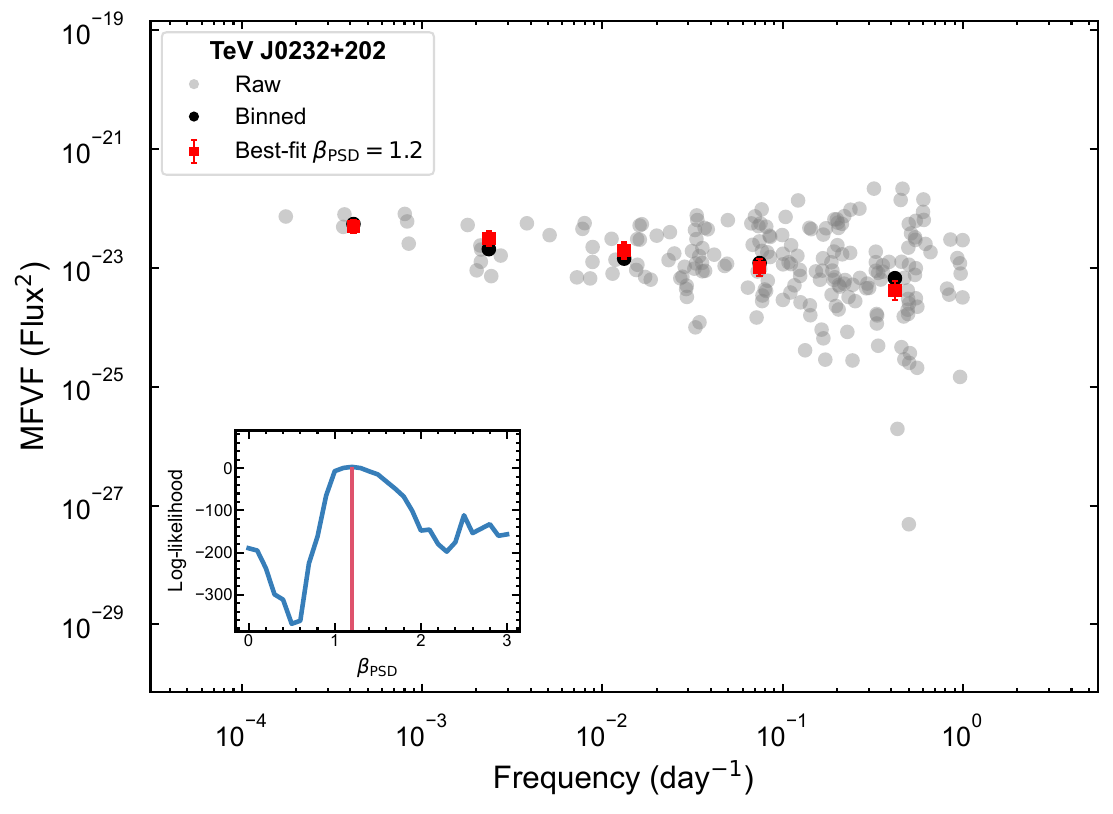}
    \caption{ 
    Variability of the source at different timescales for TeV blazars in the soft X-ray band, obtained by the multiple fragments variance function (MFVF).
    The analysis is performed only for sources with sufficiently well-sampled soft X-ray light curves. 
    Gray circles show the observed raw MFVF, and black circles show the binned observed MFVF. 
    Red squares represent the mean MFVF values in each frequency bin derived from 1000 simulated light curves based on the best-fitting PSD slope obtained through the forward-modeling procedure, while the corresponding uncertainties are derived from the standard deviation of the distribution of simulated MFVF values.
    The inset figure shows the log-likelihood as a function of PSD slope calculated by Equation~\eqref{eq:log-likelihood_PSD}, and the vertical red line marks the best-fitting PSD slope.\\
    (Only four representative sources are shown as examples.)
    }
    \label{fig:PSD_Soft}
\end{figure}

The best-fitting PSD parameters are summarized in Table~\ref{tab:PDF_fitting}, together with the total duration of each light curve. 
Figure~\ref{fig:PSD_Soft} presents representative observed MFVFs and the simulated MFVF distributions for their best-fitting PSD slopes, together with the likelihood profiles.
The resulting PSD slopes span a range of $\beta_{\rm PSD}=0.1-1.6$, with the majority of the sources having values between 1.0 and 1.6, as listed in Table~\ref{tab:PDF_fitting}.

The PSD provides a compact description of variability as a function of temporal frequency and can therefore offer qualitative clues to the underlying stochastic processes. 
In radio-quiet AGNs, X-ray PSDs commonly show red-noise behavior and are sometimes better described by bending power laws, with characteristic bend frequencies linked to black hole mass and accretion rate \citep[e.g.,][]{McHardy1987Natur, Lawrence1993ApJ, Uttley2002MNRAS, Markowitz2003ApJ, McHardy2004MNRAS, Gonzalez-Martin2012A&A}.
Blazars differ from radio-quiet AGNs because their observed X-ray emission is generally dominated by the relativistic jet. 
Their X-ray PSDs can therefore probe stochastic processes operating in the jet rather than accretion-flow variability alone. 
\citet{Mukherjee2019MNRAS486} suggested that the characteristic timescale associated with a break in the blazar PSD may reflect the typical time interval between successive shocks propagating along the jet.
Observationally, \citet{Kataoka2001ApJ560} found breaks in the PSDs of the TeV blazars Mrk~421, Mrk~501, and PKS~2155-304 at timescales of about one day or longer, and suggested that these timescales are related to the typical duration of individual flares. 
At timescales shorter than one day, the PSDs of these sources were steep, with $\beta_{\rm PSD} \sim 2-3$.
On longer timescales, flatter X-ray PSDs have been obtained for several blazars. 
For Mrk~421, \citet{Isobe2015ApJ798} obtained $\beta_{\rm PSD} = 1.6$ from three years of Monitor of All Sky X-ray Image (MAXI) monitoring. 
\citet{Goyal2020MNRAS494} found X-ray PSD slopes of $\beta_{\rm PSD} = 1.3$ for Mrk~421 and PKS~2155-304, consistent with our results of $\beta_{\rm PSD} = 1.2$ for both sources. 
A similarly broad range has been found in non-TeV blazars. 
The long-term RXTE X-ray light curve of 3C~279 obtained $\beta_{\rm PSD}=2.3$ in \citet{Chatterjee2008ApJ}. 
In contrast, \citet{Goyal2022ApJ927} obtained the flatter X-ray slopes of $\beta_{\rm PSD} = 1.2$ from the Rossi X-ray Timing Explorer (RXTE) and $\beta_{\rm PSD} = 1.4$ from Swift-XRT for the same source. 
The PSD slopes derived for our TeV-blazar sample, predominantly $1.0 \leqslant \beta_{\rm PSD} \leqslant 1.6$, are consistent with those reported from long-term X-ray monitoring of well-studied TeV blazars, suggesting that the long-term X-ray variability is driven by stochastic processes.
A few sources in our sample show relatively flat PSD slopes, with $\beta_{\rm PSD}$ = 0.1, 0.2, 0.5, 0.8, and 0.9 for TeV~J1422+323, TeV~J1031+508, TeV~J1221+282, TeV~J0904-575, and TeV~J1725+118, respectively. 
These sources have light curves with a relatively limited number of data points compared with others, as listed in Table~\ref{tab:sample}, which may affect the determination of their PSD slopes and account for the relatively flat values.

\subsubsection{Flux distribution}

A normal flux distribution is commonly associated with additive stochastic variability, whereas a log-normal distribution is typically interpreted as a signature of multiplicative variability \citep{Uttley2005MNRAS}. 
For sources in our sample with sufficiently well-sampled soft X-ray light curves, we find that the soft X-ray flux distributions are generally better described by log-normal probability density functions, with the exception of TeV~J0232+202, whose flux distribution is better described by a normal distribution (see Section~\ref{sec:fluxdist}).

The predominance of log-normal soft X-ray flux distributions is broadly consistent with previous variability studies of blazars. 
Long-term X-ray observations of BL~Lacertae provided an early clear detection of log-normal flux \citep{Giebels2009A&A503}. 
Subsequent X-ray analyses have shown that the flux distributions can be source-dependent; for example, long-term RXTE observations revealed log-normal behavior for 3C~273, whereas Mrk~421 and Mrk~501 showed more complex, potentially multi-component distributions \citep{Khatoon2020MNRAS491}. 
Log-normal behavior has also been reported at high and very-high $\gamma$-ray energies, including in long-term monitoring of PKS~2155-304 \citep{HESS2017A&A598} and in sample studies of \emph{Fermi}-LAT blazars \citep{Shah2018RAA, Wang2024ApJS270}. 
The normal distribution found for TeV~J0232+202 in our sample demonstrates that log-normal variability is not common among TeV blazars and that the variability process may depend on the source and/or its activity state.

Log-normal variability is commonly associated with multiplicative stochastic processes in accreting systems \citep{Uttley2005MNRAS}. 
In blazars, such behavior may arise if fluctuations generated in the accretion flow propagate into the relativistic jet \citep{McHardy2008bvesconfE}. 
Alternatively, multiplicative variability may be generated within the jet itself. 
The observed X-ray flux depends on several coupled physical quantities, including particle injection and acceleration, radiative cooling, and Doppler boosting, so stochastic variations in these parameters may naturally produce log-normal-like flux distributions. 
In addition, scenarios involving the superposition of multiple emitting sub-regions \citep[e.g., ``mini-jets'';][]{Biteau2012A&A} or fluctuations in characteristic particle acceleration timescales \citep{Shah2018RAA, Sinha2018MNRAS} have also been proposed to produce log-normal behavior. 
By contrast, an approximately normal flux distribution may suggest that the observed variability is dominated by the linear superposition of many statistically independent emission events or emitting regions, rather than by multiplicative processes.

\subsection{Spectral behavior}

Spectral evolution can provide insight into the balance between particle acceleration and radiative cooling in the emitting region. 
Assuming a power-law electron distribution at injection,
\begin{equation}
    \frac{\mathrm{d}N}{\mathrm{d}\gamma} = N_{0}~\gamma^{-p}, 
    \qquad \gamma_{\rm min} < \gamma < \gamma_{\rm max},
\end{equation}
where $\gamma_{\rm min}$ and $\gamma_{\rm max}$ are the minimum and maximum injected Lorentz factors, $p$ is the electron spectral index, and $N_{0}$ is a normalization related to the total particle density, the optically thin synchrotron emissivity scales as
\begin{equation}
    j_{\rm syn}(\epsilon) \propto \epsilon^{-(p-1)/2},
\end{equation}
where $\epsilon = h\nu/(m_{\rm e}c^{2})$ \citep{Chiang2002ApJ}. 
The synchrotron spectral index is therefore $\alpha_{\rm syn}=(p-1)/2$, and in the X-ray band $\alpha_{\rm syn}=\Gamma-1$ for a power-law photon spectrum. 
Because synchrotron cooling is energy dependent ($t_{\rm cool}\propto \gamma^{-1}$), changes in acceleration and cooling conditions can modify the high-energy electron population and thereby affect both the X-ray flux and photon index.

Using the flux--index correlations derived in Section~\ref{Results}, we find distinct behaviors among HSPs, ISPs, and LSPs (Figures~\ref{fig:Flux_vs_Index_Soft} and \ref{fig:Flux_vs_Index_Hard}; Table~\ref{tab:linear_regression}). 
In the discussion below, we focus on cases with $p<0.05$ as candidate correlations.

\begin{itemize}
\item \textit{HSPs.} 
All HSPs with significant correlations exhibit an anti-correlation between soft X-ray flux and photon index, indicating a harder-when-brighter trend. 
Since the soft X-ray emission of HSPs is typically dominated by the synchrotron component (Figure~\ref{fig:SED}), episodes of enhanced acceleration and/or an increased high-energy cutoff in the electron distribution can harden the X-ray spectrum and increase the flux and harden the X-ray spectrum.
This result is consistent with previous X-ray studies of TeV HSPs, in which harder spectra at higher fluxes were associated with changes in the high-energy end of electrons
\citep{Wierzcholska2016MNRAS458_56, Wang2018ApJ867, Giommi2021MNRAS}.
The absence of a significant correlation in some HSPs should not be interpreted as evidence against this interpretation, because the flux--index relation can be diluted by sparse sampling or a limited observed window.

\item \textit{LSPs.} 
TeV~J1310+323 shows an opposite trend with softer-when-brighter. 
In LSPs, X-rays often probe the rising part of the inverse Compton component and/or a regime where multiple components contribute (Figure~\ref{fig:SED}). 
An increase in the relative synchrotron contribution in the soft X-ray band (e.g., due to changes in the electron distribution) can soften the observed X-ray spectrum, producing a softer-when-brighter trend. 
We also note that TeV~J1256-057 shows harder-when-brighter behavior. 
This may reflect source-dependent component mixing (e.g., competition between non-thermal jet emission and thermal/reflective components) and/or changes in the relative dominance of spectral components across states \citep[e.g.,][]{Giommi2021MNRAS, Grandi2004Sci, Mondal2022A&A}.

\item \textit{ISPs.} 
Most ISPs show softer-when-brighter behavior. 
This trend is consistent with changes in the relative dominance of synchrotron and inverse Compton components as the source brightens \citep[e.g.,][]{Wierzcholska2016MNRAS458_56, Wierzcholska2016MNRAS458_2350}. 
Three ISPs (TeV~J1427+238, TeV~J1512-091, and TeV~J2039+523) exhibit harder-when-brighter trends. 
For TeV~J1512-091, the optical--UV accretion-disk contribution visible in the SED may bias the inferred synchrotron peak classification, and the X-rays appear consistent with sampling the rising inverse Compton component. 
\end{itemize}

\citet{Wierzcholska2016MNRAS458_56} identified X-ray spectral upturns in five TeV blazars, consistent with a superposition of the high-energy synchrotron tail and the low-energy inverse Compton component.  
Their multi-epoch analysis showed that the synchrotron component was more variable than the inverse Compton component, while the upturn energy changed modestly with flux.
They suggested that the spectral evolution could not be explained solely by changes in the cooling of the highest-energy electrons.
This result may help explain the diverse flux--index trends observed in our sample.
In HSPs, where the soft X-ray band is usually synchrotron dominated, harder-when-brighter behavior may plausibly reflect changes in particle acceleration and cooling. 
By contrast, in ISPs and LSPs, the X-ray band can cover the transition between synchrotron and inverse Compton emission, such that either harder-when-brighter or softer-when-brighter behavior may arise depending on the relative contribution of the two components.

In hard X-rays, two HSPs, TeV~J1104+382 and TeV~J1653+397, show strong anti-correlations between flux and photon index, again indicating harder-when-brighter behavior (see Figure~\ref{fig:Flux_vs_Index_Hard} and Table~\ref{tab:linear_regression}). 
This is consistent with hard X-rays tracing the high-energy synchrotron emission in these sources. 
For sources without significant correlations, limited sampling may dilute intrinsic relationships between flux and spectral shape.

The physical mechanism responsible for particle acceleration in blazar jets remains an open question. 
Among the commonly invoked scenarios, the shock-in-jet model \citep{Marscher1985ApJ298} provides a qualitative framework in which disturbances propagating along the jet can modify the energy distribution of radiating particles and thereby produce correlated flux and spectral variability. 
Such scenarios have been discussed in the context of X-ray and optical spectral evolution \citep[e.g.,][]{Rani2010MNRAS, Agarwal2015MNRAS, Xiong2017ApJS, Wang2024ApJ}, and may offer a plausible, though not unique, interpretation for the spectral trends identified in this work.

\section{Conclusion}\label{Conclusion}

We have conducted a systematic X-ray study of 85 TeV-detected blazars using archival \emph{Swift}-XRT and \emph{NuSTAR} observations.
The sample shows common X-ray variability, with 78 sources classified as variable in soft X-rays and 14 in hard X-rays.
For well-sampled soft X-ray light curves, the flux distributions are generally log-normal, except for TeV~J0232+202, and the PSD slopes are typically flat, with values of $\beta_{\rm PSD} \sim 0.1$--$1.6$.
The flux--index relations show harder-when-brighter behavior in HSPs and more diverse trends in ISPs and LSPs, consistent with changes in the dominant X-ray emission component.
We find positive correlations between TeV luminosity and both soft and hard X-ray luminosities, suggesting a physical connection between the X-ray and TeV emissions, consistent with a leptonic framework. 
The X-ray luminosity also correlates with disk luminosity, but not significantly with black hole mass, supporting an accretion–jet connection while indicating that the observed X-ray is primarily regulated by jet conditions.


\begin{deluxetable}{llcccccccccc}
\tabletypesize{\footnotesize}
\tablewidth{0pt}

\tablecaption{Sample of 85 TeV blazars
\label{tab:sample}
}
\rotate
\tablehead{
    \colhead{TeV Name} &
    \colhead{4FGL Name} &
    \colhead{$z$} &
    \colhead{$n_{\rm H,~Gal}$} &
    \colhead{$N_{\rm obs,~soft}$} &
    \colhead{${\rm Index}_{\rm soft}$} &
    \colhead{$F_{\rm soft}$} &
    \colhead{$F_{\rm var,~soft}$} &
    \colhead{$N_{\rm obs,~hard}$} &
    \colhead{${\rm Index}_{\rm hard}$} &
    \colhead{$F_{\rm hard}$} &
    \colhead{$F_{\rm var,~hard}$}
}
\colnumbers
\startdata
    J0013-188 & J0013.9-1854 & 0.09485 & $1.97 \times 10^{-2}$ & 10 & $1.86 \pm 0.09$ & $(1.33 \pm 0.12) \times 10^{-11}$ & $0.183 \pm 0.032$ & -- & -- & -- & -- \\
    J0033-193 & J0033.5-1921 & 0.61 & $1.52 \times 10^{-2}$ & 6 & $2.24 \pm 0.07$ & $(9.14 \pm 0.48) \times 10^{-12}$ & $0.445 \pm 0.023$ & -- & -- & -- & -- \\
    J0035+598 & J0035.9+5950 & 0.086 & $3.94 \times 10^{-1}$ & 157 & $0.47 \pm 0.16$ & $(1.46 \pm 0.28) \times 10^{-10}$ & $0.646 \pm 0.018$ & 1 & $2.49 \pm 0.02$ & $(2.12 \pm 0.13) \times 10^{-11}$ & -- \\
    J0112+227 & J0112.1+2245 & 0.265 & $3.74 \times 10^{-2}$ & 20 & $2.22 \pm 0.10$ & $(6.76 \pm 0.45) \times 10^{-12}$ & $0.761 \pm 0.016$ & 2 & $1.61 \pm 0.03$ & $(6.67 \pm 0.71) \times 10^{-12}$ & N/A$^{a}$ \\
    J0136+391 & J0136.5+3906 & 0.75 & $6.86 \times 10^{-2}$ & 10 & $2.12 \pm 0.06$ & $(3.69 \pm 0.20) \times 10^{-11}$ & $0.144 \pm 0.017$ & -- & -- & -- & -- \\
    ... & ... & ... & ... & ... & ... & ... & ... & ... & ... & ... & ... \\
\enddata
\tablecomments{
Col. (1): the TeV blazar name;
Col. (2): the 4FGL name;
Col. (3): the source redshift;
Col. (4): the Galactic hydrogen column density in units of $\times 10^{22} \rm cm^{-2}$ from \citet{HI4PI2016};
Col. (5): the number of soft X-ray observations;
Col. (6): the arithmetic mean photon index of soft X-ray observations;
Col. (7): the arithmetic mean flux of soft X-ray observations in units of $\rm erg ~ cm^{-2} ~ s^{-1}$;
Col. (8): the fractional variability of soft X-ray observations;
Col. (9): the number of hard X-ray observations;
Col. (10): the arithmetic mean photon index of hard X-ray observations;
Col. (11): the arithmetic mean flux of hard X-ray observations in units of $\rm erg ~ cm^{-2} ~ s^{-1}$;
Col. (12): the fractional variability of hard X-ray observations.\
\tablenotetext{a}{The fractional variability is not reported because the numerator in Equation~\eqref{Eq:F_var} is negative, indicating that the observed variability is weak relative to the measurement uncertainties.
}
(Only five rows are listed here.)
}
\end{deluxetable}


\begin{deluxetable}{lcccc}
\tabletypesize{\footnotesize} 
\tablewidth{0pt}
\tablecaption{Linear regression and Pearson correlation results
\label{tab:OLS_Pearson}
}
\tablehead{
\colhead{Relation} &
\colhead{$k$} &
\colhead{$b$} &
\colhead{$r$} &
\colhead{$p$}
}
\colnumbers
\startdata
    $\log~L_{\rm TeV}$ vs $\log~L_{\rm X,~Soft}$ & $1.54 \pm 0.20$ & $-24.89 \pm 9.07$ & 0.63 & $< 10^{-4}$ \\
    $\log~L_{\rm TeV}$ vs $\log~L_{\rm X,~Hard}$ & $1.23 \pm 0.28$ & $-10.99 \pm 12.67$ & 0.52 & $1.66 \times 10^{-3}$ \\
    \midrule
    $\log$ $M_{\rm BH}$ vs $\log~L_{\rm X,~Soft}$ & $-1.02 \pm 0.06$ & $54.94 \pm 2.73$ & -0.22 & 0.21 \\
    $\log~L_{\rm disk}$ vs $\log~L_{\rm X,~Soft}$ & $1.50 \pm 0.18$ & $-23.88 \pm 7.92$ & 0.61 & $< 10^{-4}$ \\
    $\log$ $M_{\rm BH}$ vs $\log~L_{\rm X,~Hard}$ & $0.98 \pm 0.07$ & $-35.50 \pm 3.24$ & 0.08 & 0.77 \\
    $\log~L_{\rm disk}$ vs $\log~L_{\rm X,~Hard}$ & $1.41 \pm 0.19$ & $-19.93 \pm 8.47$ & 0.77 & $2.38 \times 10^{-4}$ \\
\enddata
\tablecomments{
Col. (1): the relation;
Col. (2): the slope and its uncertainty;
Col. (3): the intercept and its uncertainty;
Col. (4): the Pearson correlation coefficient;
Col. (5): the Pearson chance probability.
The linear regression is described as: $y = k x + b $, where $y$ and $x$ are the variables, $k$ and $b$ are the slope and intercept, respectively. 
}
\end{deluxetable}

\begin{deluxetable}{lcccccccc}
\tabletypesize{\footnotesize}
\tablewidth{2pt} 
\tablecaption{Fitting parameters of soft X-ray flux distributions and power spectral density slopes for TeV blazars with well-sampled light curves
\label{tab:PDF_fitting}
}
\tablehead{ 
    \multirow{2}[2]{*}{TeV Name} & \multicolumn{3}{c}{Normal} & \multicolumn{3}{c}{Log-normal} & \multirow{2}[2]{*}{$\beta_{\rm PSD}$} & \multirow{2}[2]{*}{$T_{\rm PSD}$} \\
    \cmidrule(lr){2-4} \cmidrule(lr){5-7}
          & $\mu$ & $\sigma$ & $\chi^{2}_{\rm red}$ & $\mu_{\log}$ & $\sigma_{\log}$ & $\chi^{2}_{\rm red}$ & &
}
\colnumbers 
\startdata
    J0035+598 & 145.83 & 99.51 & 10.95 & 4.77 & 0.67 & 1.35 & 1.40 & 6487.52 \\
    J0214+517 & 17.25 & 3.94 & 1.81 & 2.82 & 0.22 & 1.26 & 1.10 & 5491.63 \\
    J0222+430 & 5.56 & 3.41 & $1.13\times10^{5}$ & 1.61 & 0.41 & 42.00 & 1.10 & 5592.69 \\
    J0232+202 & 25.92 & 8.52 & 1.23 & 3.19 & 0.36 & 2.53 & 1.20 & 5695.75 \\
    J0348-278 & 4.01 & 1.66 & 1.41 & 1.31 & 0.40 & 0.38 & 1.00 & 1378.24 \\
    J0416+010 & 30.56 & 10.31 & 3.58 & 3.37 & 0.33 & 1.57 & 1.30 & 5519.23 \\
    J0509+056 & 3.63 & 1.83 & 46.22 & 1.20 & 0.40 & 6.62 & 1.10 & 5281.77 \\
    J0521+211 & 22.40 & 20.07 & 16.94 & 2.76 & 0.83 & 2.98 & 1.10 & 5184.79 \\
    J0550-322 & 67.98 & 16.00 & 2.39 & 4.19 & 0.24 & 1.40 & 1.10 & 7026.99 \\
    J0650+250 & 84.10 & 42.05 & 3.24 & 4.29 & 0.56 & 2.22 & 1.40 & 4916.07 \\
    J0710+591 & 48.17 & 18.19 & 9.95 & 3.81 & 0.37 & 8.16 & 1.30 & 5532.80 \\
    J0721+713 & 20.76 & 16.28 & 115.67 & 2.78 & 0.70 & 2.39 & 1.30 & 6514.46 \\
    J0854+201 & 10.21 & 6.43 & $7.24\times10^{6}$ & 2.19 & 0.49 & 5.84 & 1.30 & 6914.66 \\
    J0904-575 & 7.28 & 2.46 & 3.56 & 1.93 & 0.32 & 1.96 & 0.80 & 2682.51 \\
    J0958+655 & 10.97 & 7.08 & 4.65 & 2.20 & 0.64 & 1.52 & 1.20 & 6428.02 \\
    J1015+494 & 48.23 & 40.89 & 10.51 & 3.51 & 0.90 & 2.14 & 1.60 & 6701.96 \\
    J1031+508 & 40.37 & 13.85 & 2.87 & 3.64 & 0.35 & 1.19 & 0.20 & 5760.83 \\
    J1103-234 & 79.79 & 16.29 & 3.20 & 4.36 & 0.20 & 2.42 & 1.20 & 6831.19 \\
    J1104+382 & 982.50 & 659.85 & $1.08\times10^{7}$ & 6.67 & 0.70 & 1.97 & 1.20 & 6983.96 \\
    J1136+701 & 28.59 & 34.32 & 19.81 & 2.75 & 1.10 & 2.78 & 1.40 & 5319.45 \\
    J1159+292 & 5.42 & 2.09 & 6.02 & 1.60 & 0.45 & 2.27 & 1.20 & 5741.68 \\
    J1221+282 & 4.58 & 3.48 & 81.93 & 1.34 & 0.56 & 3.86 & 0.50 & 3085.06 \\
    J1221+301 & 76.08 & 45.54 & 113.95 & 4.16 & 0.60 & 1.74 & 1.20 & 6483.06 \\
    J1224+213 & 6.08 & 2.38 & 7.30 & 1.74 & 0.35 & 2.72 & 1.10 & 4045.92 \\
    J1256-057 & 19.97 & 12.97 & $1.88\times10^{10}$ & 2.87 & 0.46 & 46.73 & 1.20 & 6671.25 \\
    J1310+323 & 4.45 & 1.91 & 11.61 & 1.41 & 0.41 & 0.79 & 1.20 & 6223.61 \\
    J1422+323 & 4.71 & 1.66 & 2.61 & 1.49 & 0.35 & 2.04 & 0.10 & 1906.35 \\
    J1427+238 & 15.70 & 14.24 & 21.36 & 2.43 & 0.75 & 6.66 & 1.40 & 4676.22 \\
    J1428+426 & 60.10 & 16.39 & 3.60 & 4.06 & 0.27 & 1.14 & 1.20 & 6815.60 \\
    J1512-091 & 12.89 & 3.53 & 4.66 & 2.52 & 0.27 & 1.83 & 1.10 & 6192.80 \\
    J1555+111 & 75.04 & 47.89 & 14.09 & 4.10 & 0.70 & 2.65 & 1.40 & 6941.88 \\
    J1653+397 & 206.93 & 127.01 & 127.34 & 5.16 & 0.60 & 3.18 & 1.40 & 6993.36 \\
    J1725+118 & 13.56 & 8.52 & 12.60 & 2.47 & 0.49 & 2.96 & 0.90 & 5627.65 \\
    J1728+502 & 52.81 & 35.18 & 15.65 & 3.75 & 0.67 & 3.46 & 1.50 & 4806.24 \\
    J1751+096 & 8.79 & 5.33 & 3.96 & 2.00 & 0.58 & 1.50 & 1.30 & 5994.63 \\
    J1959+651 & 453.42 & 237.70 & 31.91 & 5.98 & 0.55 & 1.24 & 1.30 & 6900.35 \\
    J2009-488 & 54.04 & 72.38 & 31.72 & 3.58 & 0.75 & 13.03 & 1.60 & 5825.59 \\
    J2039+523 & 12.41 & 7.42 & 4.23 & 2.33 & 0.64 & 2.05 & 1.30 & 1095.05 \\
    J2158-302 & 92.89 & 63.67 & 103.28 & 4.32 & 0.67 & 0.94 & 1.20 & 6594.40 \\
    J2202+422 & 27.35 & 25.38 & $3.43\times10^{3}$ & 3.09 & 0.62 & 381.34 & 1.20 & 6730.69 \\
    J2347+517 & 29.59 & 14.37 & $9.28\times10^{3}$ & 3.29 & 0.42 & 3.04 & 1.00 & 6842.01 \\
    J2359-306 & 37.50 & 15.67 & 3.86 & 3.54 & 0.41 & 2.36 & 1.60 & 3082.59 \\
\enddata
\tablecomments{ 
Col. (1): the TeV blazar name; 
Col. (2): the mean value of the normal distribution;
Col. (3): the standard deviation of the normal distribution;
Col. (4): the reduced $\chi^{2}$ value of the normal distribution;
Col. (5): the mean value of the log-normal distribution;
Col. (6): the standard deviation of the log-normal distribution;
Col. (7): the reduced $\chi^{2}$ value of the log-normal distribution;
Col. (8): the PSD slope;
Col. (9): the total duration, in days, of the light curve over which the PSD slope is obtained.
}
\end{deluxetable}


\begin{deluxetable}{lcccccccc}
\tabletypesize{\footnotesize}
\tablewidth{2pt} 
\tablecaption{Slopes and intercepts of linear regressions between photon index and X-ray flux for TeV blazars with well-sampled X-ray light curves
\label{tab:linear_regression}
}
\tablehead{ 
    \multirow{2}[0]{*}{TeV Name} & \multicolumn{4}{c}{Soft X-rays} & \multicolumn{4}{c}{Hard X-rays} \\
    \cmidrule(lr){2-5} \cmidrule(lr){6-9}
          & $r$ & $p$ & $k$ & $b$ & $r$ & $p$ & $k$ & $b$
}
\colnumbers 
\startdata
J0035+598 & -0.43 & $<1\times10^{-4}$ & -0.91 $\pm$ 0.06 & -8.54 $\pm$ 0.61 & -- & -- & -- & -- \\
J0214+517 & -0.38 & $3.15\times10^{-3}$ & -1.57 $\pm$ 0.21 & -15.24 $\pm$ 2.26 & -- & -- & -- & -- \\
J0222+430 & 0.44 & $<1\times10^{-4}$ & 1.16 $\pm$ 0.12 & 15.48 $\pm$ 1.32 & -- & -- & -- & -- \\
J0232+202 & -0.11 & $1.17\times10^{-1}$ & -1.02 $\pm$ 0.02 & -9.23 $\pm$ 0.22 & -- & -- & -- & -- \\
J0348-278 & -0.30 & $6.09\times10^{-2}$ & -1.04 $\pm$ 0.09 & -10.22 $\pm$ 0.99 & -- & -- & -- & -- \\
J0416+010 & -0.28 & $4.63\times10^{-2}$ & -1.01 $\pm$ 0.07 & -8.36 $\pm$ 0.74 & -- & -- & -- & -- \\
J0509+056 & 0.47 & $1.40\times10^{-4}$ & 1.66 $\pm$ 0.18 & 20.96 $\pm$ 2.07 & -0.47 & $5.08\times10^{-2}$ & -0.78 $\pm$ 0.17 & -7.18 $\pm$ 1.94 \\
J0521+211 & -0.29 & $6.44\times10^{-3}$ & -1.14 $\pm$ 0.07 & -11.15 $\pm$ 0.76 & -- & -- & -- & -- \\
J0550-322 & -0.32 & $2.23\times10^{-3}$ & -1.19 $\pm$ 0.11 & -10.39 $\pm$ 1.08 & -- & -- & -- & -- \\
J0650+250 & -0.16 & $1.21\times10^{-1}$ & -0.89 $\pm$ 0.06 & -7.06 $\pm$ 0.65 & -- & -- & -- & -- \\
J0710+591 & -0.39 & $4.84\times10^{-4}$ & -0.82 $\pm$ 0.07 & -6.68 $\pm$ 0.77 & -- & -- & -- & -- \\
J0721+713 & 0.38 & $<1\times10^{-4}$ & 0.95 $\pm$ 0.03 & 12.43 $\pm$ 0.33 & -- & -- & -- & -- \\
J0854+201 & 0.50 & $<1\times10^{-4}$ & 1.35 $\pm$ 0.03 & 16.84 $\pm$ 0.34 & -- & -- & -- & -- \\
J0904-575 & 0.03 & $8.67\times10^{-1}$ & 1.04 $\pm$ 0.13 & 13.11 $\pm$ 1.45 & -- & -- & -- & -- \\
J0958+655 & -0.17 & $9.16\times10^{-2}$ & -1.01 $\pm$ 0.04 & -9.77 $\pm$ 0.42 & -- & -- & -- & -- \\
J1015+494 & -0.74 & $<1\times10^{-4}$ & -0.57 $\pm$ 0.05 & -3.73 $\pm$ 0.48 & -- & -- & -- & -- \\
J1031+508 & -0.07 & $7.14\times10^{-1}$ & -0.95 $\pm$ 0.11 & -7.88 $\pm$ 1.13 & -- & -- & -- & -- \\
J1103-234 & -0.05 & $6.91\times10^{-1}$ & -1.04 $\pm$ 0.09 & -8.59 $\pm$ 0.94 & -- & -- & -- & -- \\
J1104+382 & -0.86 & $<1\times10^{-4}$ & -0.85 $\pm$ 0.01 & -5.57 $\pm$ 0.11 & -0.87 & $<1\times10^{-4}$ & -0.53 $\pm$ 0.04 & -2.36 $\pm$ 0.42 \\
J1136+701 & -0.81 & $<1\times10^{-4}$ & -0.57 $\pm$ 0.04 & -3.84 $\pm$ 0.46 & -- & -- & -- & -- \\
J1159+292 & -0.02 & $8.54\times10^{-1}$ & -1.01 $\pm$ 0.04 & -9.78 $\pm$ 0.46 & -- & -- & -- & -- \\
J1221+282 & 0.04 & $8.08\times10^{-1}$ & 1.02 $\pm$ 0.04 & 14.01 $\pm$ 0.50 & -- & -- & -- & -- \\
J1221+301 & -0.59 & $<1\times10^{-4}$ & -0.70 $\pm$ 0.05 & -5.14 $\pm$ 0.51 & -- & -- & -- & -- \\
J1224+213 & 0.46 & $<1\times10^{-4}$ & 1.73 $\pm$ 0.15 & 21.11 $\pm$ 1.69 & -- & -- & -- & -- \\
J1256-057 & -0.57 & $<1\times10^{-4}$ & -0.75 $\pm$ 0.03 & -6.49 $\pm$ 0.37 & -- & -- & -- & -- \\
J1310+323 & 0.53 & $<1\times10^{-4}$ & 1.36 $\pm$ 0.15 & 17.22 $\pm$ 1.66 & -- & -- & -- & -- \\
J1422+323 & 0.36 & $3.20\times10^{-2}$ & 1.41 $\pm$ 0.24 & 17.62 $\pm$ 2.68 & -- & -- & -- & -- \\
J1427+238 & -0.47 & $<1\times10^{-4}$ & -0.67 $\pm$ 0.07 & -4.87 $\pm$ 0.72 & -- & -- & -- & -- \\
J1428+426 & -0.25 & $1.88\times10^{-4}$ & -0.98 $\pm$ 0.03 & -8.14 $\pm$ 0.32 & -- & -- & -- & -- \\
J1512-091 & -0.28 & $<1\times10^{-4}$ & -1.21 $\pm$ 0.06 & -11.84 $\pm$ 0.64 & -- & -- & -- & -- \\
J1555+111 & -0.57 & $<1\times10^{-4}$ & -0.57 $\pm$ 0.02 & -3.58 $\pm$ 0.21 & -- & -- & -- & -- \\
J1653+397 & -0.84 & $<1\times10^{-4}$ & -0.84 $\pm$ 0.01 & -6.16 $\pm$ 0.12 & -0.93 & $<1\times10^{-4}$ & -0.53 $\pm$ 0.03 & -3.05 $\pm$ 0.32 \\
J1725+118 & 0.05 & $7.79\times10^{-1}$ & 1.00 $\pm$ 0.02 & 13.01 $\pm$ 0.21 & -- & -- & -- & -- \\
J1728+502 & -0.65 & $<1\times10^{-4}$ & -0.66 $\pm$ 0.03 & -4.87 $\pm$ 0.33 & -- & -- & -- & -- \\
J1751+096 & 0.05 & $7.84\times10^{-1}$ & 1.01 $\pm$ 0.03 & 12.69 $\pm$ 0.38 & -- & -- & -- & -- \\
J1959+651 & -0.54 & $<1\times10^{-4}$ & -0.78 $\pm$ 0.02 & -5.53 $\pm$ 0.21 & -- & -- & -- & -- \\
J2009-488 & -0.27 & $3.96\times10^{-2}$ & -0.94 $\pm$ 0.09 & -7.46 $\pm$ 0.97 & -- & -- & -- & -- \\
J2039+523 & -0.49 & $3.43\times10^{-3}$ & -1.91 $\pm$ 0.32 & -20.08 $\pm$ 3.49 & -- & -- & -- & -- \\
J2158-302 & -0.43 & $<1\times10^{-4}$ & -0.66 $\pm$ 0.03 & -4.14 $\pm$ 0.26 & -0.09 & $7.96\times10^{-1}$ & -1.00 $\pm$ 0.04 & -8.00 $\pm$ 0.38 \\
J2202+422 & 0.07 & $7.41\times10^{-2}$ & 1.05 $\pm$ 0.03 & 12.56 $\pm$ 0.34 & -- & -- & -- & -- \\
J2347+517 & -0.31 & $<1\times10^{-4}$ & -1.08 $\pm$ 0.04 & -9.68 $\pm$ 0.40 & -- & -- & -- & -- \\
J2359-306 & -0.27 & $1.53\times10^{-1}$ & -0.94 $\pm$ 0.08 & -7.81 $\pm$ 0.89 & -- & -- & -- & -- \\
\enddata
\tablecomments{ 
The linear regression is described as $y = kx + b$, where $y$ and $x$ are the variables, and $k$ and $b$ are the slope and intercept, respectively.
Symbol ``--'' represents the null value.
}
\end{deluxetable}


\appendix

\section{Tables of X-ray Observations}
\label{Appendix:Xray_Data}

The X-ray measurements used in this work are provided in machine-readable form. 
Because the Swift-XRT and NuSTAR data have different instrumental quantities, we provide separate tables for the two instruments. 
The Swift-XRT table includes the observation information, count rate, spectral parameters, unabsorbed 0.3--10 keV flux, fit statistic, readout mode, and derived luminosity. 
The NuSTAR table includes the FPMA/FPMB exposures and count rates, cross-normalization factor, spectral parameters, unabsorbed 3--60 keV flux, fit statistic, and derived luminosity. 
Examples of the table formats are shown in Tables~\ref{tab:xrt_observation_example} and \ref{tab:nustar_observation_example}.

\begin{deluxetable*}{lccccccccc}
\tabletypesize{\scriptsize}
\setlength{\tabcolsep}{2.5pt}
\tablecaption{Example of the Swift-XRT Observation Table
\label{tab:xrt_observation_example}}
\tablehead{
\colhead{Source} &
\colhead{ObsID} &
\colhead{Exposure} &
\colhead{MJD} &
\colhead{Count Rate} &
\colhead{$\Gamma$ ($\alpha$)} &
\colhead{$\beta$} &
\colhead{$F_{\rm 0.3-10\,keV}$} &
\colhead{C-stat/DoF} &
\colhead{Mode}
}
\startdata
    TeV J0013-188 & 00031806002 & 3800.86 & 55449.12 &
    $0.2436 \pm 0.0080$ &
    $1.83_{-0.06}^{+0.06}$ &
    $0.39_{-0.15}^{+0.15}$ &
    $1.43_{-0.13}^{+0.13}\times10^{-11}$ &
    291.12/303 & PC \\
    TeV J0013-188 & 00031806003 & 4215.40 & 55450.09 &
    $0.2864 \pm 0.0083$ &
    $1.93_{-0.04}^{+0.04}$ &
    -- &
    $1.57_{-0.06}^{+0.06}\times10^{-11}$ &
    309.43/333 & PC \\
    TeV J0013-188 & 00031806004 & 3858.30 & 55451.12 &
    $0.2573 \pm 0.0082$ &
    $1.83_{-0.06}^{+0.06}$ &
    $0.48_{-0.14}^{+0.15}$ &
    $1.36_{-0.12}^{+0.12}\times10^{-11}$ &
    288.66/304 & PC \\
    TeV J0013-188 & 00031806005 & 2744.99 & 55452.08 &
    $0.2605 \pm 0.0098$ &
    $1.96_{-0.06}^{+0.06}$ &
    -- &
    $1.39_{-0.07}^{+0.07}\times10^{-11}$ &
    252.51/281 & PC \\
    TeV J0013-188 & 00031806006 & 1525.85 & 56925.33 &
    $0.1321 \pm 0.0094$ &
    $1.44_{-0.16}^{+0.16}$ &
    $1.03_{-0.34}^{+0.36}$ &
    $1.23_{-0.25}^{+0.26}\times10^{-11}$ &
    116.28/135 & PC \\
    ... & ... & ... & ... &
    ... &
    ... &
    ... &
    ... &
    ... & ... \\
\enddata
\tablecomments{
Only the first five rows are shown here. The complete Swift-XRT table is available in machine-readable form. Exposure is given in seconds, MJD is the effective observing time, count rate is in counts s$^{-1}$, and flux is in erg cm$^{-2}$ s$^{-1}$. Symbol ``--'' indicates that the parameter is not available.
}
\end{deluxetable*}

\begin{deluxetable*}{lccccccccccc}
\tabletypesize{\scriptsize}
\setlength{\tabcolsep}{1.8pt}
\tablecaption{Example of the NuSTAR Observation Table
\label{tab:nustar_observation_example}}
\tablehead{
\colhead{Source} &
\colhead{ObsID} &
\colhead{Exposure A} &
\colhead{Exposure B} &
\colhead{MJD} &
\colhead{Count Rate A} &
\colhead{Count Rate B} &
\colhead{${\rm Const}_{\rm B/A}$} &
\colhead{$\Gamma$ ($\alpha$)} &
\colhead{$\beta$} &
\colhead{$F_{\rm 3-60\,keV}$} &
\colhead{C-stat/DoF}
}
\startdata
    TeV J0035+598 & 60160017002 & 23279.71 & 23049.70 & 59998.35 &
    $0.2498 \pm 0.0033$ &
    $0.2382 \pm 0.0033$ &
    $1.0096_{-0.0191}^{+0.0195}$ &
    $2.49_{-0.02}^{+0.02}$ &
    -- &
    $2.12_{-0.13}^{+0.13}\times10^{-11}$ &
    1083.53/1081 \\
    TeV J0112+227 & 60801029002 & 57315.95 & 56277.54 & 59905.67 &
    $0.0288 \pm 0.0007$ &
    $0.0274 \pm 0.0007$ &
    $0.9936_{-0.0363}^{+0.0376}$ &
    $1.59_{-0.03}^{+0.03}$ &
    -- &
    $6.45_{-0.80}^{+0.83}\times10^{-11}$ &
    986.27/1030 \\
    TeV J0112+227 & 60801029004 & 100473.60 & 99547.95 & 59908.18 &
    $0.0324 \pm 0.0006$ &
    $0.0311 \pm 0.0006$ &
    $1.0017_{-0.0260}^{+0.0266}$ &
    $1.63_{-0.02}^{+0.02}$ &
    -- &
    $6.88_{-0.60}^{+0.61}\times10^{-12}$ &
    1344.44/1331 \\
    TeV J0214+517 & 60160094002 & 21439.83 & 21364.04 & 57417.61 &
    $0.1718 \pm 0.0029$ &
    $0.1596 \pm 0.0028$ &
    $0.9439_{-0.0227}^{+0.0232}$ &
    $2.05_{-0.02}^{+0.02}$ &
    -- &
    $1.78_{-0.14}^{+0.14}\times10^{-11}$ &
    1082.26/1128 \\
    TeV J0222+430 & 60601022002 & 24608.71 & 24358.41 & 59143.25 &
    $0.0187 \pm 0.0009$ &
    $0.0183 \pm 0.0009$ &
    $1.0444_{-0.0706}^{+0.0757}$ &
    $1.82_{-0.07}^{+0.07}$ &
    -- &
    $3.41_{-0.80}^{+0.85}\times10^{-12}$ &
    476.85/558 \\
    ... & ... & ... & ... & ... &
    ... &
    ... &
    ... &
    ... &
    ... &
    ... &
    ... \\
\enddata
\tablecomments{
Only the first five rows are shown here. 
The complete NuSTAR table is available in machine-readable form. 
Exposure A and Exposure B are the FPMA and FPMB exposures in seconds, Count Rate A and Count Rate B are the corresponding net count rates in counts s$^{-1}$, ${\rm Const}_{\rm B/A}$ is the cross-normalization factor of FPMB relative to FPMA, and $F_{\rm 3-60\,keV}$ is the unabsorbed 3--60 keV flux in erg cm$^{-2}$ s$^{-1}$. 
Symbol ``--'' represents the null value. 
}
\end{deluxetable*}


\begin{acknowledgments}
We sincerely thank the anonymous reviewer for the careful review and constructive comments, which helped improve the quality and clarity of the manuscript.
H.B.X. acknowledges the support from the National Natural Science Foundation of China (NSFC, grant No. 12203034), the Shanghai Science and Technology Fund (grant No. 22YF1431500), the China Manned Space Project (grant No. CMS-CSST-2025-A07), and the Shanghai Municipal Education Commission regarding artificial intelligence-empowered research.
This work is supported by the National Natural Science Foundation of China (NSFC) grant 12261141691.
M.M. acknowledges the Croatian Science Foundation (HrZZ) under the project number IP-2022-10-4595 and the University of Rijeka Project uniri-mzi-25-3 funded by the European Union - NextGenerationEU.
JHFan’s work is partially supported by the National Natural Science Foundation of China (grant Nos.12433004) and the National Key Research and Development Program of China, Grant No. 2025YFA1614102.
The \textit{Swift}-XRT data are supplied by the UK Swift Science Data Centre at the University of Leicester.
\end{acknowledgments}


\bibliography{Ref}{}
\bibliographystyle{aasjournal}



\end{document}